\documentclass[11pt,letterpaper]{article}

\usepackage{jheppub}
\usepackage{bigstrut}
\usepackage{multirow}
\usepackage{booktabs}
\usepackage{amssymb}
\usepackage{dsfont}
\usepackage{epsfig,multicol,bbm}
\usepackage{amsfonts}
\usepackage{amsmath}
\usepackage{mqcommands}

\newcommand{\reef}[1]{(\ref{#1})}
\renewcommand{\eqref}[1]{(\ref{#1})}

\title{Bootstrapping Conformal Field Theories with the Extremal Functional Method}

\author[a]{Sheer El-Showk,}
\author[b]{Miguel F. Paulos}
\affiliation[a]{Institut de Physique Th\'eorique CEA Saclay, CNRS-URA 2306, 91191 Gif sur Yvette, France}
\affiliation[b]{Department of Physics, Brown University, Box 1843, Providence, RI 02912-1843, USA}

\abstract{
The existence of a positive linear functional acting on the space of (differences between) conformal blocks has been shown to rule out regions in the parameter space of conformal field theories (CFTs). We argue that at the boundary of the allowed region the extremal functional contains, in principle, enough information to determine the dimensions and OPE coefficients of an infinite number of operators appearing in the correlator under analysis. Based on this idea we develop the Extremal Functional Method (EFM), a numerical procedure for deriving the spectrum and OPE coefficients of CFTs lying on the boundary (of solution space). We test the EFM by using it to rederive the low lying spectrum and OPE coefficients of the 2d Ising model based solely on the dimension of a single scalar quasi-primary -- no Virasoro algebra required. Our work serves as a benchmark for applications to more interesting, less known CFTs in the near future.
}


\begin{document}
\maketitle

\section{Introduction \& Preliminaries}
It is well known \cite{Polyakov:1970xd,
Mack:1969rr,Ferrara:1971vh,Ferrara:1973vz,Ferrara:1974nf,Ferrara:1974ny,Ferrara:1974pt,YellowBook} that the global conformal symmetry group in $D$ dimensions $SO(D+1,1)$ implies that the four-point function of four identical scalars with conformal dimension $\Delta_\sigma$ takes the form
	\bea
	\langle \sigma (x_1)\sigma (x_2)\sigma (x_3)\sigma (x_4)\rangle=\frac{g(u,v)}{x_{12}^{2\Delta_\sigma}\,x_{34}^{2\Delta_\sigma}},
	\eea
with $x_{ij}\equiv x_i-x_j$, and where $g(u,v)$ is a function of the conformally invariant cross-ratios
	\bea
	u=\frac{x_{12}^2\, x_{34}^2}{x_{13}^2\, x_{24}^2}, \quad v=\frac{x_{14}^2\,x_{23}^2}{x_{13}^2\,x_{24}^2}.
	\eea
The existence of the operator product expansion (OPE) of the theory \cite{Wilson:1969zs} implies that the function $g(u,v)$ can be written in two inequivalent ways, corresponding to expanding the correlator around $x_1\simeq x_2,x_3\simeq x_4$ or $x_1\simeq x_3, x_2\simeq x_4$ (i.e. the direct and crossed channels, respectively). For instance, in the direct channel we have
	\bea
	g(u,v)=1+\sum_{\Delta,L} \lambda_{\mathcal O_{\Delta,L}}^2 \, G_{\Delta,L}(u,v). \label{crossing}
	\eea
The sum is over symmetric traceless spin-$L$ highest-weight representations of the conformal group labelled by the conformal dimension $\Delta$ -- in other words, over towers of conformal primaries\footnote{These are primaries of the global conformal group. In two-dimensions they are usually referred to as quasi-primaries to distinguish them from Virasoro primaries. In this context, they are operators annihilated by $L_1$.} $\mathcal O$ and all their descendants. The contribution from each such tower is given by the functions $G_{\Delta,L}(u,v)$, known as conformal blocks or conformal partial waves, which are eigenfunctions of the Casimir of the conformal group\footnote{The literature on conformal blocks has been steadily growing in recent years. A partial list of the most recent and/or useful results is given by \cite{DO1,DO2,DO3,Costa:2011dw,SimmonsDuffin:2012uy}}. The accompanying numbers, $\lambda_{\mathcal O_{\Delta,L}}$, are the OPE coefficients appearing in the three-point function $\langle \sigma \sigma \mathcal O_{\Delta,L}\rangle$.   The 1 in the sum above is the contribution of the conformal block of the identity which always appears in the four-point function of identical scalars.

Equivalence of the expansion in the direct and crossed channels implies the non-trivial identity 
\bea\label{crossing1} 
v^{\Delta_\sigma} g(u,v) = u^{\Delta_\sigma} g(v,u). 
\eea
This constraint is known as a crossing relation, and a function $g(u,v)$ satisfying it, together with a conformal block expansion such as \reef{crossing} is said to be crossing symmetric. Finding crossing symmetric functions is very non-trivial: for instance the contribution of a single conformal block in one channel must necessarily be matched by an infinite sum of conformal blocks in the cross channel. The idea of using relation \reef{crossing1}, together with analogous ones for other correlation functions, to determine the spectrum and OPE coefficients of a CFT is known as the conformal bootstrap \cite{Ferrara:1973yt,Polyakov:1974gs}.

\begin{figure}
	\centering
\includegraphics[scale=0.8]{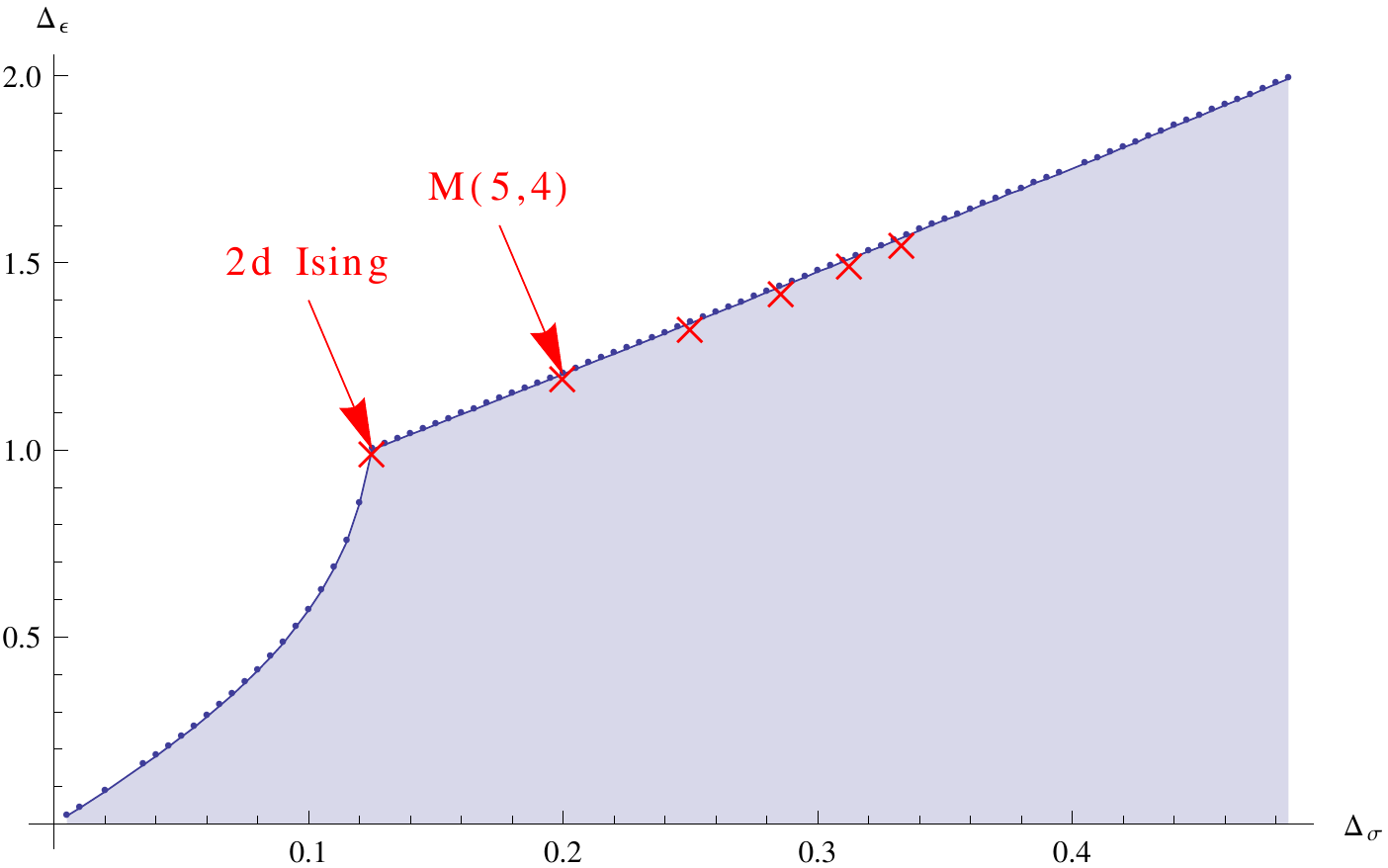}
	\centering
\caption{The plot above depicts a crossing symmetry bounds plot.  The shaded blue region corresponds to values of ($\Delta_\sigma, \Delta_\epsilon$) consistent with crossing symmetry.  Note here $\Delta_\epsilon$ is {\em defined} as the first scalar appearing in the $\sigma \, \sigma$ OPE.  Note the kink at the value $\Delta_\epsilon \approx 1.000003$ corresponding to the two-dimensional Ising model. Some other minimal models are marked with crosses (in red).}
\label{fig:2dScan}
\end{figure}

Since crossing symmetry is really a functional statement there is a continuously infinite set of constraints on the continuously infinite set of parameters $\lambda_{\mathcal O_{\Delta,L}}$. Up until recently it was not known how to extract useful information from this equation. This all changed with the seminal work of \cite{Rattazzi:2008pe}, soon followed by \cite{Rychkov:2009ij,Caracciolo:2009bx,Rattazzi:2010gj,Poland:2010wg,Rattazzi:2010yc,Poland:2011ey,Vichi:2011ux,ElShowk:2012ht,Liendo:2012hy}. In these works it was shown that, assuming unitarity, the crossing relations impose $\Delta_\sigma$-dependent bounds on the allowed value of the first scalar appearing in the $\sigma \sigma$ OPE, for any unitary conformal field theory.  
For instance, in two dimensions \cite{Rattazzi:2008pe,Rychkov:2009ij} this bound is reproduced in figure \ref{fig:2dScan}.
%
There are several interesting features in this plot. One of them is that the minimal models nearly saturate the bound. The second is that there is an apparent kink close to $(\Delta_\sigma,\Delta_\epsilon)=(\frac 18,1)$, which we recognize as the dimension of two (Virasoro) primary operators in the $D=2$ Ising model. One of our results will be to show that the Ising model indeed saturates the bound to extremely high accuracy, although we will not investigate the issue of the existence (or not) of the kink.

\subsection{Bootstrapping via Crossing Symmetry}

To see how this kind of constraints can be derived, we start by rephrasing equation \reef{crossing1} as
	\bea
	\sum_{\Delta,L} \lambda_{\mathcal O}^2\, \left(\frac{v^{\Delta_\sigma}\, G_{\Delta,L}(u,v)-u^{\Delta_\sigma}\, G_{\Delta,L}(v,u)}{u^{\Delta_\sigma}-v^{\Delta_\sigma}}\right)\equiv \sum_{\Delta,L} \lambda_{\mathcal O}^2\, F^{(\sigma)}_{\Delta,L}(u,v)=1 \label{cross2}
	\eea
where we have selected out the contribution of the conformal block of the identity (i.e. one) and left it on the RHS.  It is useful to think of the $F^{(\sigma)}_{\Delta,L}(u,v)$ as a continuous set of vectors labelled by $\Delta, L$ (but depending also on $\Delta_\sigma$ which is kept fixed).  These functions are vectors in the formal sense of being elements of the infinite dimensional vector space of functions (on the plane) but in practice we expand these functions in a power series around a point to some finite order reducing the problem to a finite dimensional one.
For more details on this procedure, and a more complete exposition of other points above, such as  the definitions of conformal blocks in diverse dimensions, the reader is referred to \cite{Rattazzi:2008pe, ElShowk:2012ht}.

Unitarity immediately reduces the size of our vector space as not all possible values of $\Delta,L$ are allowed. Rather, only $\Delta$ satisfying
	\bea
	\Delta\geq \frac{D-2}{2}, \quad L=0; \\
	\Delta\geq D-L+2, \quad L>0,
	\eea
are compatible with unitarity.
Other than this requirement, a given crossing symmetric function could include any number of operators with arbitrary spin and dimension, and the sum above should really be understood as an integral in the generic case. What really matters is the set of OPE coefficients, which can of course contain zeroes meaning certain operators do not appear in the correlator. 

The crucial handle into attacking the problem is to notice the positivity of the coefficients $\lambda_{\mathcal O_{\Delta,L}}^2$ (following from unitarity), which leads us to consider all possible positive linear combinations of the vectors $F^{(\sigma)}_{\Delta,L}$. These combinations form a semi-polyhedral cone, i.e. roughly a cone which has smooth ``faces'' along certain directions, and polyhedral faces along others. So, we can rephrase the problem of solving the bootstrap constraints \reef{cross2} as the question: under which assumptions does the cone generated by the $F^{(\sigma)}_{\Delta,L}$ vectors contain the constant function $1$, the ``identity vector''? 
This kind of problem has been well studied before, and falls into the class of Linear Programming problems, i.e. a minimization problem given some set of linear inequalities (see e.g. \cite{linearprog}). There are efficient numerical algorithms for solving such problems, such as the simplex method, suitable for implementation on a computer.  In fact our problem is even simpler as we do not need to solve a minimization problem but only to check whether a particular vector lies within a semi-polyhedral cone defined by a set of inequalities.

The next step is to derive the bound proper. One way of doing this is to fix $\Delta_\sigma$ and impose successively stronger constraints on the spectrum of operators. We can for instance demand that in the sum \reef{cross2} we shall not include any scalars which have conformal dimension below some cutoff. Now, initially, for low cutoffs, one finds there is always a solution to crossing symmetry but as we increase the cutoff the shape of the cone changes. At some point the identity vector will pass through a face of the cone; if we increase the cutoff further the identity vector will lie outside the cone, and it will no longer be possible to find a solution to the crossing symmetry constraints (consistent with the truncated spectrum). We can also phrase this as the fact that in the vector space under consideration it is now possible to find a hyperplane which neatly separates the vectors inside the cone from the vector corresponding to the identity.  The existence of such a hyperplane corresponds to the existence of a ``dual vector'', $\phi$, which we will refer to as a linear functional, which has a positive inner product with all the $F_{\Delta, L}$ but a negative inner product with the identity vector.  Since our vector space is constructed from derivatives of the $F^{(\sigma)}_{\Delta,L}$ the dual vector $\phi$ is nothing other than a derivative operator.  The existence of such an operator or linear functional will play an essential role in what follows so we provide a much more detailed exposition in section \ref{sec:warmup}.

\subsection{The Extremal Functional}

From the description above we see that something very interesting happens when the identity vector lies on the face of the polytope formed by the $F^{(\sigma)}_{\Delta,L}$.  At precisely this point there is generically a single solution to crossing symmetry. That is because at this point we are only allowed to use the vectors which make up the vertices of the face through which the identity vector is cutting across. Indeed, by convexity, if we try adding any other vector, we will move irremediably away from the face containing the identity vector. In the language of linear functionals, the hyperplane contains the face of the cone along which  the identity vector lies. By definition then, the zeroes of this unique functional are precisely the vectors which define the face through which the identity vector is passing.\footnote{The fact that the zeroes of the functional provide a solution to crossing symmetry was already mentioned in \cite{Poland:2010wg}, although its uniqueness (and therefore relevance) was not emphasized.
}

To summarize, at the boundary of the allowed region, the spectrum of operators which are allowed in possible solutions of crossing symmetry is severely constrained: one can only include the vectors which are zeroes of the linear functional. Furthermore, generically one finds this solution is unique, since otherwise there would be extra linear dependencies among vectors defining the face. This uniqueness provides a recipe for constructing solutions to crossing symmetry, the {\it Extremal Functional Method} (EFM):

\begin{itemize}
\item Find the extremal linear functional $\phi$.
\item Compute the vectors $F^{(\sigma)}_{\Delta,L}$ which are zeroes of $\phi$.
\item Solve for the linear combination of $F^{(\sigma)}_{\Delta,L}$'s which gives the identity vector. The coefficients are the square of the OPE coefficients.
\end{itemize}
The goal of this short note is to show that this program is not only feasible, but that it yields highly accurate results for the low-lying spectrum of a CFT. We will focus on the test case of the two-dimensional Ising model. We do this for two reasons: first, we shall see that this model saturates the bound curve and so is directly accessible using our methods; second, the spectrum of primaries and OPE coefficients is known, giving us a benchmark by which to fine-tune our numerical methods. Our sole dynamical input will be the dimension of $\sigma$. Given this number, we shall show that our method fixes the dimensions and OPE coefficients of a significant number of operators to very high accuracy. Indeed, from the discussion above, we see that in principle EFM completely fixes an infinite set of operators. In practice, we are limited by computational power and numerical precision, but we clearly see a steady increase in the number of accurate operator dimensions and OPE coefficients as we increase the dimension of our vector space. 

In this first investigation we shall show how with limited computer power -- a laptop and a few hours of running time -- we obtain about fifteen OPE coefficients to within $1\%$ accuracy (and a further 4 to within 10\%).  Of these, six coefficients are accurate to within $0.01\%$  including the stress-tensor contribution, $\lambda_{\sigma\sigma T}$, from which we derive the central charge to 6 digits precision. The story for the operator dimensions is better yet and there is still vast room for technical improvement, on which work is currently underway.  We are quite confident that this will push the number of accurate operators into the several dozens.  While these results are obviously far weaker than exact two-dimensional methods it should be noted that in the three-dimensional Ising models, where we expect these techniques to apply, essentially no OPE coefficients are known (and only a handful of operator dimensions).

Here is a short outline of this note. We begin with a pedagogical introduction to the EFM by working with a lowest order approximation. In section \ref{sec:Ising} we describe in detail the application of our method to the two-dimensional Ising model. We show that the Ising model saturates the dimension bound to extremely high accuracy. From the extremal functional we read off the spectrum of the theory. We show that as the dimension of the vector space is increased the number of zeroes both increases and stabilizes at the expected theoretical values. This is important for future applications of our method to e.g. the $D=3$ Ising model as there we will not have the theoretical results to guide us, and hence it is important to define a convergence criterion which will tell us how much we should trust the spectrum obtained. We discuss this in detail and determine the spectrum and OPE coefficients with our method, finding excellent agreement with the theoretical predictions. We finish with a discussion of the limitations of our procedure and future improvements. In the appendices we discuss technical details of our numerical methods and give detailed plots and tables of our results. 
In the online version of this note we include a notebook with more detailed technical information.

\section{Warm-up: 2 derivatives.}\label{sec:warmup}

Let us see how EFM works in its simplest setting. The first thing to do in order to obtain a tractable problem is to discretize the infinite set of constraints in \reef{cross2}. We do this by first setting $u=z\bar z, v=(1-z)(1-\bar z)$, followed by an expansion around $z=\bar z=1/2$. More concretely we set $z=\frac 12 (1+a+\sqrt{b}), \bar z=\frac 12(1+a-\sqrt{b})$ and expand in $a$ and $b$. Effectively we are choosing a basis for the infinite set of constraints given by its derivatives around $u=v=1/4$, a basis which has been shown to lead to good results in previous studies \cite{Rattazzi:2008pe}. After discretization the constraints \reef{crossing} can be thought of as demanding that the constant vector $1$ lies in the cone spanned by positive linear combinations of the vectors $F^{(\sigma)}_{\Delta,L}$. 

We begin with the simplest non-trivial example where we consider the constant term and linear terms in $a$ and $b$ in the expansion, giving us three component vectors. That is, equation \reef{cross2} now becomes the vector equation
\bea\label{crossVect}
	\sum_{\Delta,L} \lambda_{\mathcal O_{\Delta,L}}^2 
	\left(
	\begin{array}{c}
	F^{(\sigma)}_{\Delta,L}(\frac 14,\frac 14)\\
	\partial_a F^{(\sigma)}_{\Delta,L}(\frac 14,\frac 14)\\
	\partial_b F^{(\sigma)}_{\Delta,L}(\frac 14,\frac 14)
	\end{array}
	\right)=\left(\begin{array}{c}
	1\\
	0\\
	0
	\end{array}
	\right)
	\eea
We are free to redefine the coefficients $\lambda_{\mathcal O_{\Delta,L}}^2$ to $\hat \lambda_{\mathcal O_{\Delta,L}}^2=\lambda_{\mathcal O_{\Delta,L}}^2 F^{(\sigma)}_{\Delta,L}(\frac 14,\frac 14)$.  As we are initially simply interested in establishing bounds on operator dimensions this rescaling is innocuous but we must of course rescale back once we wish to compute OPE coefficients. The first equation has become the normalization condition
	\bea
	\sum_{\Delta,L} \hat \lambda_{\mathcal O_{\Delta,L}}^2=1
	\eea
This reduces the dimension of the vector space by one and we are now looking at a slice of the cone corresponding to rescaling the top component of all vectors to one (i.e. the intersection of the original cone with the plane transverse to $(1, 0, 0)$). Such a slice yields a two-dimensional convex polytope, which is nothing but the convex hull of all possible vectors $(\partial_a F/F,\partial_b F/F)$ and we are interested in determining the circumstances under which the origin is contained in it.
\begin{centering}
	\begin{figure}
		\begin{tabular}{@{\hskip -0.4cm}c@{\hskip -0.7cm}c@{\hskip -0.7cm}c}
	\includegraphics[scale=1]{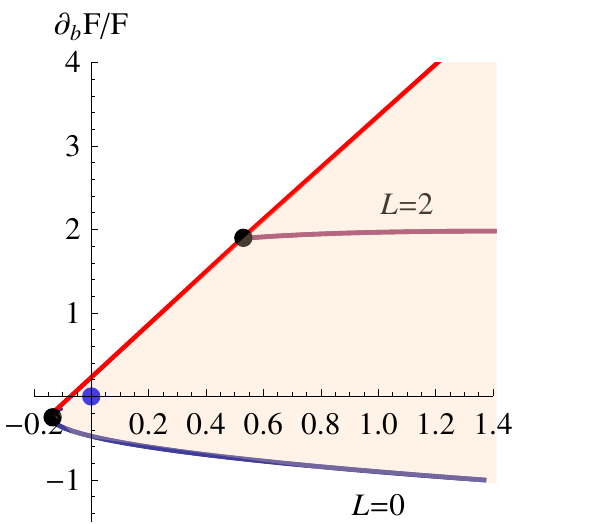}
	&
	\includegraphics[scale=1]{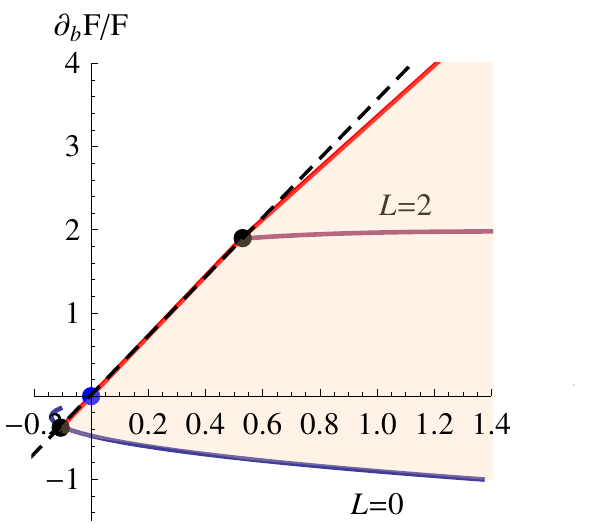}
	&
	\includegraphics[scale=1]{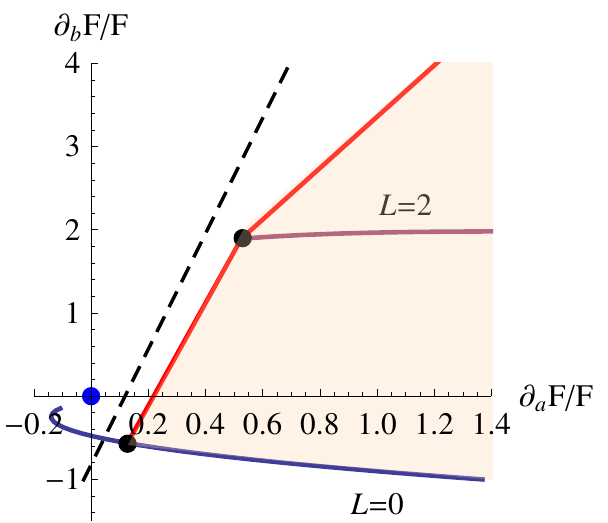}\\
	(a)&(b)&(c)
\end{tabular}
\caption{Convex hull formed by the $F^{(\sigma)}_{\Delta,L}$ vectors. In blue and purple the spin-0 and spin-2 vectors, respectively.  The lines start at the unitarity bounds ($\Delta=0,2$ respectively on the left and $\Delta$ increases to the right). Higher spin curves lie outside the plot. The red curve shows part of the boundary of the polytope, the other being the spin-0 line in blue. A solution to crossing symmetry exists whenever the origin (in blue) is contained inside the polytope. In the critical case (b) a solution must involve the vectors marked with black dots. The dashed line is the linear functional separating the origin from the polytope, and it overlaps with a polytope face in the critical case. }
\label{fig:2dIsing2der}
\end{figure}
\end{centering}

Figure \ref{fig:2dIsing2der} shows the cone in $(\partial_a F, \partial_b F)$ space defined by all possible values of $\Delta, L$ consistent with unitarity and for  $\Delta_{\sigma}=0.125$. Represented are the scalar and spin-2 lines, corresponding to the sets of vectors with spins $0$ and $2$, and arbitrary $\Delta$. Increasing the conformal dimension shifts the vectors to the right-hand-side of the plot, tracing the lines shown there in blue and purple (i.e. each point along the colored lines corresponds to a given $\Delta$ with $L=0,2$, respectively). There are also other lines corresponding to higher spin fields lying outside the range of the plot above. Altogether, the convex hull of all these vectors forms an unbounded polytope, one of whose faces runs from the tips of the scalar and spin-2 lines, another being the scalar line in blue.

Now we implement the procedure described in the introduction. If we allow all possible spin-0 fields, it is clear that the origin lies well inside the polytope (fig \ref{fig:2dIsing2der}(a)). This means that within our approximation, we cannot rule out CFTs with this unrestricted spectrum, which is good since we know free theories certainly exist. As we start imposing a gap on the scalar spectrum (i.e. require all scalars to satisfy $\Delta > \Delta_\epsilon$ for some $\Delta_\epsilon$), the boundary of the polytope shifts, following the scalar line up to a critical point $\Delta_{\epsilon}^*$. At this critical point the face which runs from the tip of the scalar line to the tip of the spin-2 line cuts across the origin (fig \ref{fig:2dIsing2der}(b)). Finally in figure fig.\ref{fig:2dIsing2der}(c) we see that the origin is not contained in the polytope and hence there is no solution to the crossing constraints. Accordingly it is possible to find a linear functional separating the identity vector from the polytope, as shown by the dashed line in the same figure.

There are two special things about the critical point in fig \ref{fig:2dIsing2der}(b)). First, if we pass it, there is no longer a solution to crossing symmetry, since the origin will no longer be in the convex hull of all the other vectors (fig \ref{fig:2dIsing2der}(c)). Secondly, precisely at this critical point the solution is unique. By convexity it is clear that the solution to crossing symmetry must include the critical scalar of dimension $\Delta_{\epsilon}^*$, and the field  at the tip of the spin-2 line, which is of course nothing but the stress-tensor, and no other operator. So, at this point we have found the two unique operators which must be in the spectrum within this two-derivative approximation. Numerically one finds that the critical scalar has dimension $\Delta_{\epsilon}\simeq 1.03$.

To compute the OPE coefficients we simply solve \reef{crossVect} for $\lambda_{ {\cal O}_{{\Delta_\epsilon},0}}$ and $\lambda_{ {\cal O}_{2,2}}$ (with all other $\lambda$ vanishing), yielding a system of two equations for two variables, whose solution is:
	\bea
	\Delta_\sigma=0.125, &&\qquad \Delta_\epsilon\simeq 1.03 \nonumber \\
	\lambda_{\sigma \sigma \epsilon}\simeq 0.24, &&\qquad c\simeq 0.45
	\eea
with $c$ the central charge, related to the $\sigma \sigma T$ OPE coefficient $\lambda_{\sigma \sigma T}$ by $c=\Delta_\sigma^2/\lambda_{\sigma \sigma T}^2$.
Recall the exact values should be
	\bea
	\Delta_\sigma=0.125, \qquad \Delta_\epsilon= 1 \nonumber \\
	\lambda_{\sigma \sigma \epsilon}= 0.25, \qquad c= 0.5
	\eea
and so even with this very basic two derivative approximation we already get something quite reasonable.

\begin{centering}
	\begin{figure}
\begin{tabular}{cc}
	\includegraphics[scale=0.7]{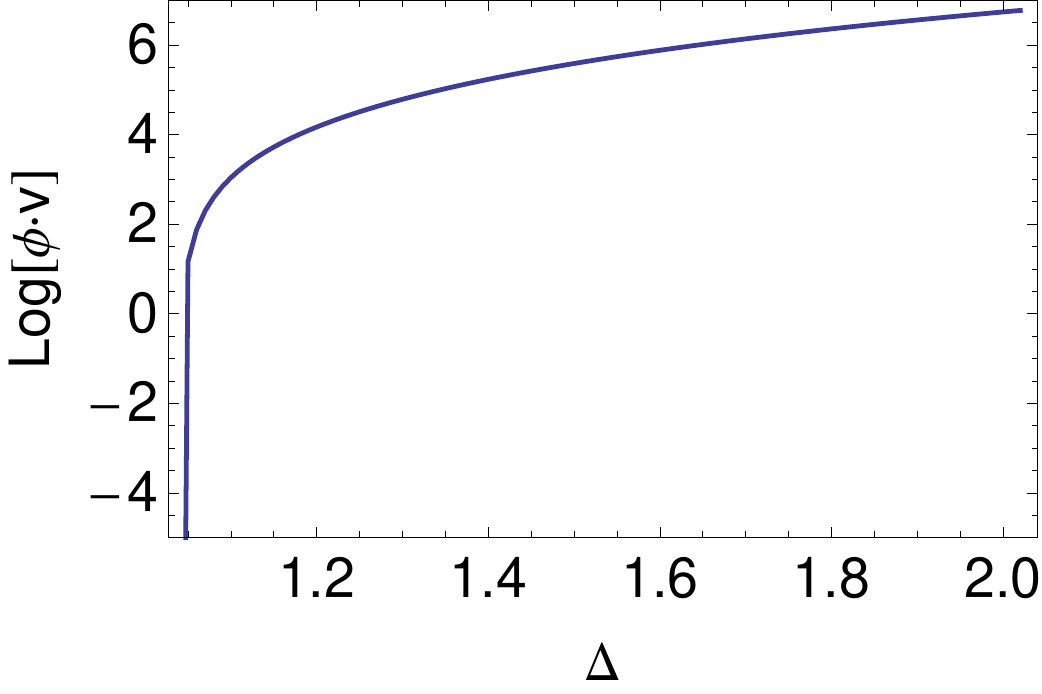}
	&
	\includegraphics[scale=0.73]{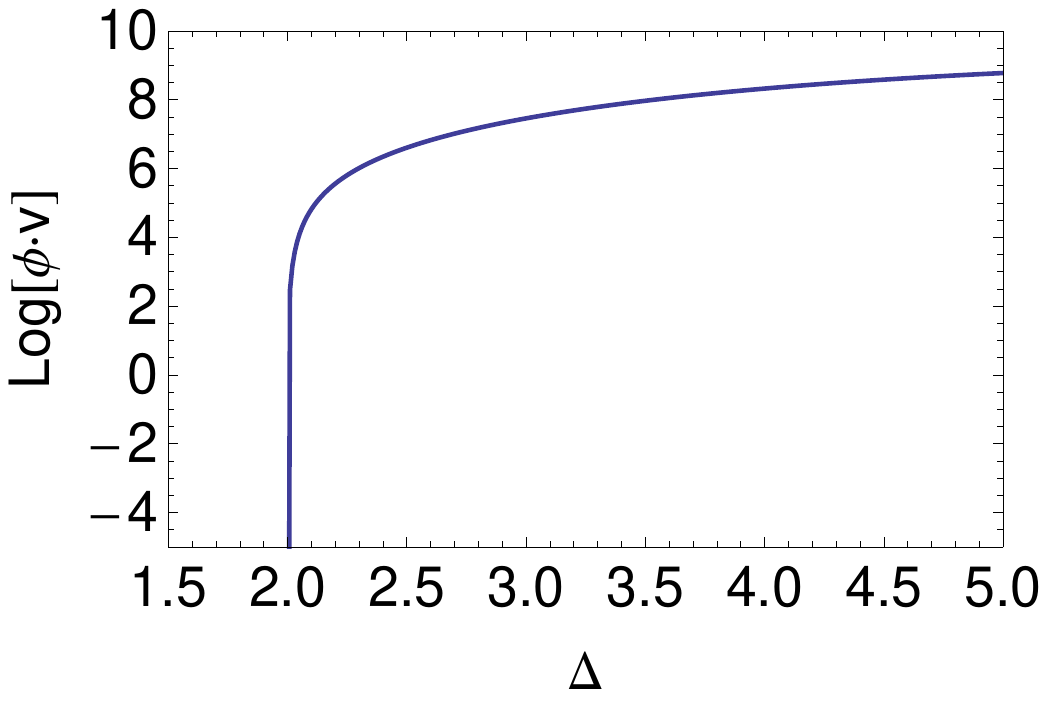}
	\\	
	(a)&(b)
\end{tabular}
\caption{The extremal functional $\phi$ acting on the spin-0 (a), and spin-2 (b) vectors. Notice the logarithmic scale. There are two clear zeroes, corresponding to operators $(1.03,0)$ and $(2,2)$. The functional is positive when acting on all other vectors.}
\label{fig:2dIsingfunc}
\end{figure}
\end{centering}

It is useful to take a different perspective on the preceding calculation. Instead of asking when can we solve the crossing symmetry relations, we can ask when can we {\it not} solve them. This takes us to the language of hyperplanes and linear functionals that we mentioned in the introduction. Here the idea is that if we want to guarantee that the identity vector is not contained inside the polyhedral cone, then we need to find a hyperplane, (in this case just an ordinary plane), which separates the identity from all other vectors. In practice we can demand that a linear functional exists which is positive when acting on all vectors except the identity, where it should be constrained to be negative. If we can find it, then we have proven that there is no solution to crossing symmetry.
In the polytope picture, the plane becomes a line (i.e. the intersection of the plane with the plane transverse to (1,0,0)). In fig. \ref{fig:2dIsing2der}(b) the critical line overlaps with the face spanned by the critical scalar and the stress-tensor. These two vectors are therefore zeroes of the original linear functional, and the functional should be positive when acting on all other vectors. This behaviour is shown in fig \ref{fig:2dIsingfunc}.

\section{Applying EFM to the $D=2$ Ising Model}
\label{sec:Ising}

Our simple example in the previous section shows all the features that we discussed in the introduction: at the boundary of the allowed region in the $(\Delta_\sigma,\Delta_\epsilon)$ plane the linear functional has a set of zeroes, being positive everywhere else; the zeroes of the functional are the only operators which we should include to solve the constraints of crossing symmetry; and solving such constraints at a given order in derivatives fixes completely the respective OPE coefficients. It is with this linear functional approach that we shall proceed to analyse more complicated cases involving more derivatives. We include some of the technical details in the appendix; see also \cite{ElShowk:2012ht}. Essentially, we add more derivatives of the $F$ functions with respect to $a,b$, making the vectors larger, and thereby enlarging the dimension of our search space. In this way we are capturing more and more information about the shape of the full cone, and therefore about the spectrum. A natural (and it turns out good) way to parameterize this approach is the number, $N$, of components in our vectors or simply the dimension of the search space. 

To select out the 2d Ising model we set $\Delta_\sigma=0.125$ and look for an extremal functional.  That is we fix $\Delta_\sigma$ and find the maximal possible value of the next scalar operator $\Delta_\epsilon$ (i.e. the largest possible gap in the scalar spectrum).  We emphasize that $\Delta_\epsilon$ is not being fixed by hand but by maximizing the gap in the scalar spectrum.

\begin{figure}[t]
	\centering
\includegraphics[scale=0.5]{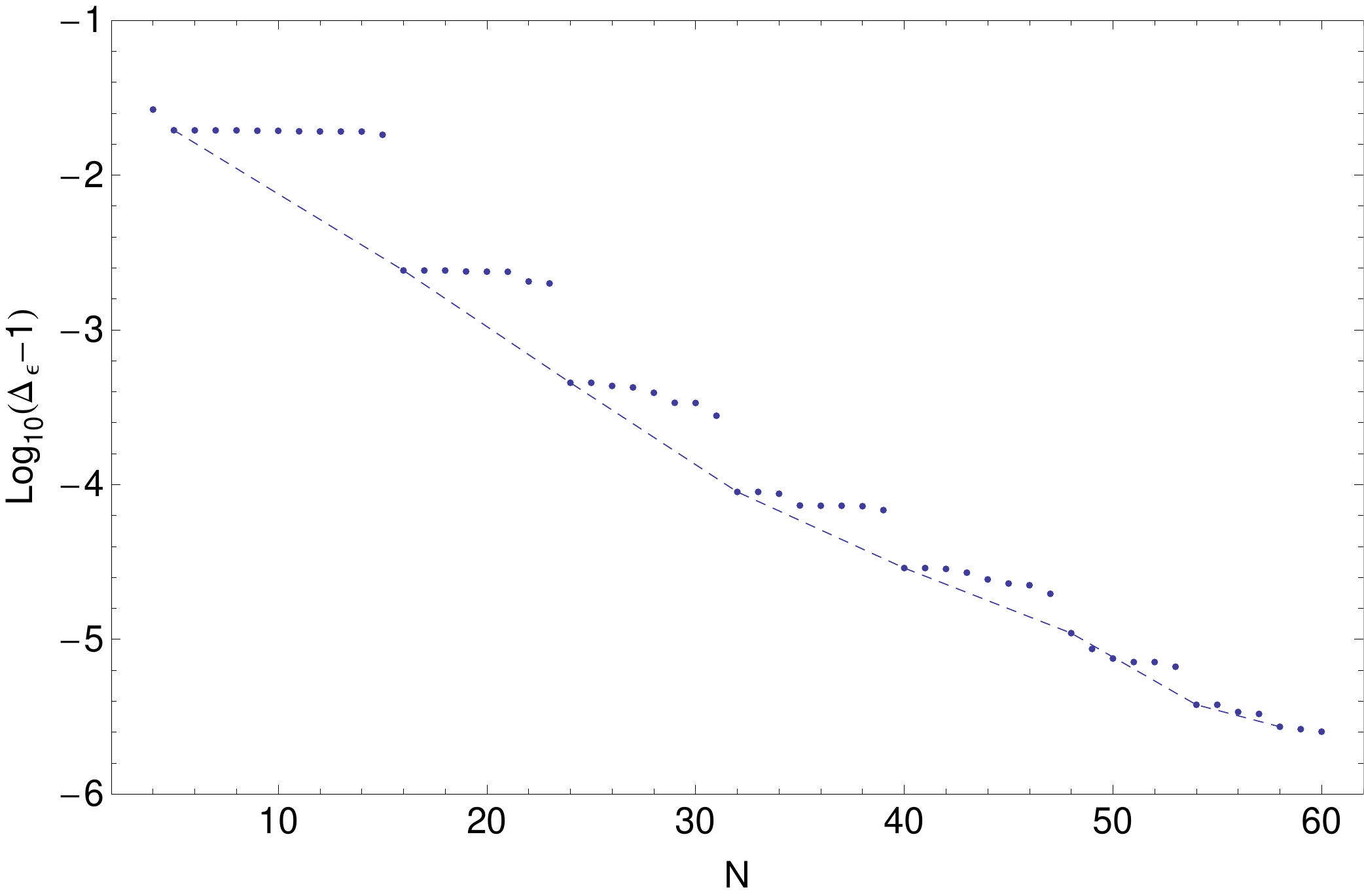}
	\caption{Evolution of the critical dimension $\Delta_{\epsilon}$ as the number of derivatives is increased.}
	\label{fig:epsIsing}
\end{figure}

We start with small $N$ and gradually increase it up to a maximum of 60 components. As the number of derivatives is increased, the boundary of the allowed region shrinks towards lower values of $\Delta_{\epsilon}^*$. We are guaranteed to improve our bound as $N$ increases since, for a given $N$, we can always use the functional at $N-1$ by simply adding an extra zero component.  We find empirically that indeed the value of $\Delta_\epsilon^*$ systematically decreases towards $\Delta_{\epsilon}^*=1$. This behaviour can be clearly seen in the plot in figure \ref{fig:epsIsing}.
For the maximum value of $N$ we considered here we obtain the correct value $\Delta_{\epsilon}=1$ to about six decimal places. The discrete jump structure is an artifact of our organization of the search space. Indeed, a careful examination of the points at which such jumps occurs shows that they are completely correlated with the inclusion of more derivatives along the $z=\bar z$ direction. Whenever this happens there is a sudden improvement in the bound, followed by flatter behaviour as we add further transverse derivatives. In this sense, this shows that the longitudinal direction is more important for determining the spectrum than the transverse one. In any case, if we focus on these ``jumping'' points only, it becomes clear that the convergence is exponentially fast, as evidenced by the approximately linear nature of the curve connecting these points in the logarithmic plot in fig. \ref{fig:epsIsing}.

These special jump points will be important for us, since they provide useful reference points at which to evaluate the evolution of our data. Since our results vary quite a bit between jumps, and not so much in between them, whenever we shall consider issues of convergence we will always be making comparisons between successive jumping points, and not between $N$, $N+1$. For a detailed specification of the structure of the search space and a determination of these points we refer the reader to appendix \ref{lp}.

\subsection{The Spectrum} 

\begin{figure}
	\centering
\includegraphics[scale=0.5]{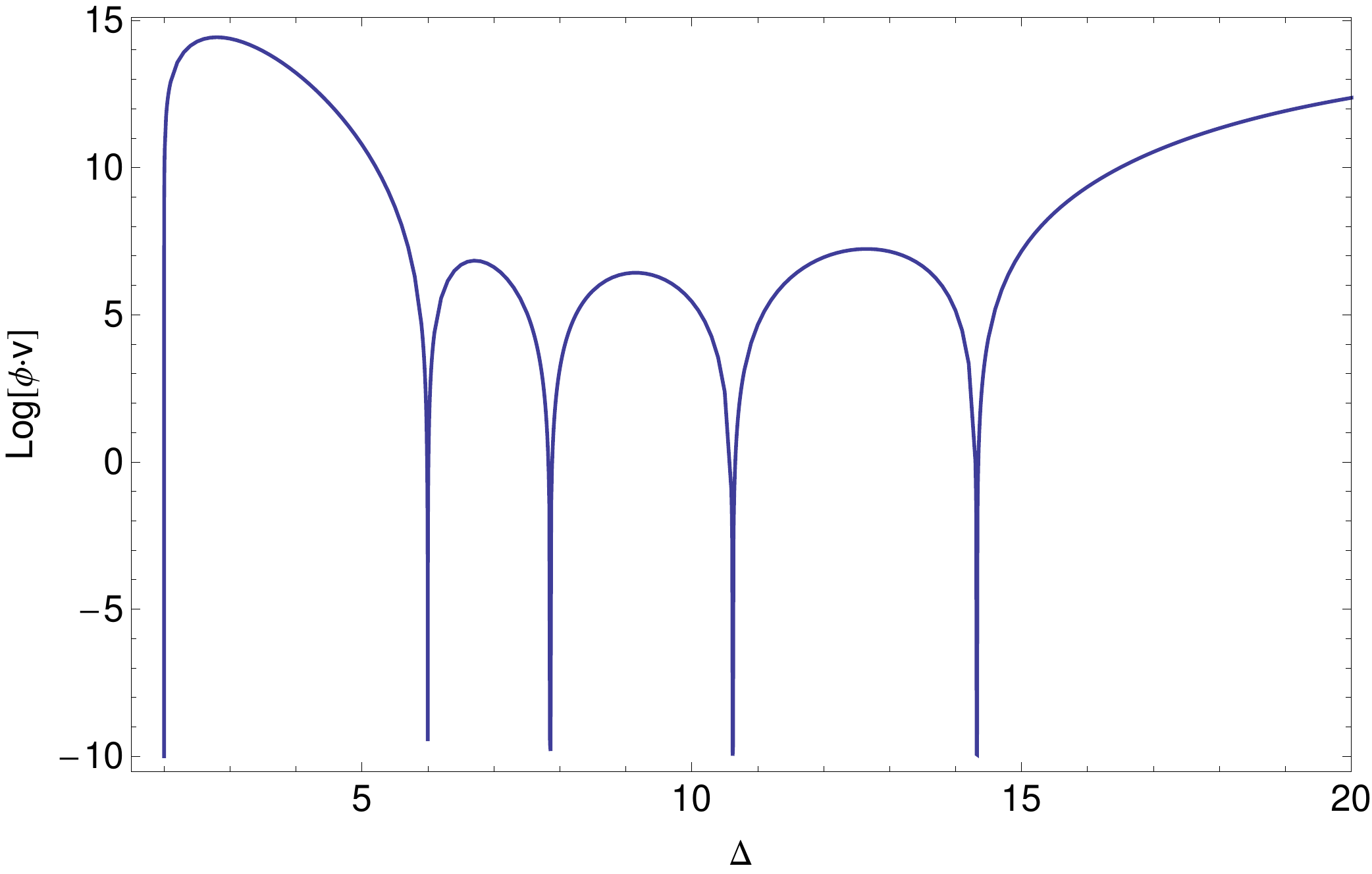}
\caption{The extremal functional acting on the spin-2 vectors.}
\label{fig:IsingFunctionalsSpin0}
\end{figure}

We now act with the extremal functional $\phi$ on the vectors $v\equiv V_{\Delta,L}$ and plot the result as a function of $\Delta$ for a given spin $L$. The value of $\phi\cdot v$ tells us roughly how far the vectors are from the hyperplane. In figure \ref{fig:IsingFunctionalsSpin0} we show a plot of the extremal functional dotted into the conformal block vectors of spin 2; see also table \ref{fig:IsingFunctionals} in appendix \ref{plots} for a more complete set of plots.
Notice the logarithmic scale. The sharp downward spikes correspond to the zeroes of the functional, so for instance from this spin-2 plot we read off operators with dimensions $\simeq 2,6,7.9,\ldots$. We will compare these spectra with those of the 2d Ising model soon enough. But first we need to address an important question, which is: which operators should we trust to be in the true spectrum? Indeed, much as $\Delta_\epsilon$ converges to the correct value only for a high enough number of derivatives, the same is true for the other operators. We therefore need a criterion for deciding whether a given operator has stabilized or not. 

Firstly, it is easy to read off the zeroes of the extremal functional and see how they behave for increasing $N$. We show such plots for spins 0 through $10$ in appendix \ref{plots} in figure \ref{fig:IsingOperators}. Here we reproduce only the scalar plot.
\begin{figure}
	\centering
\includegraphics[scale=0.6]{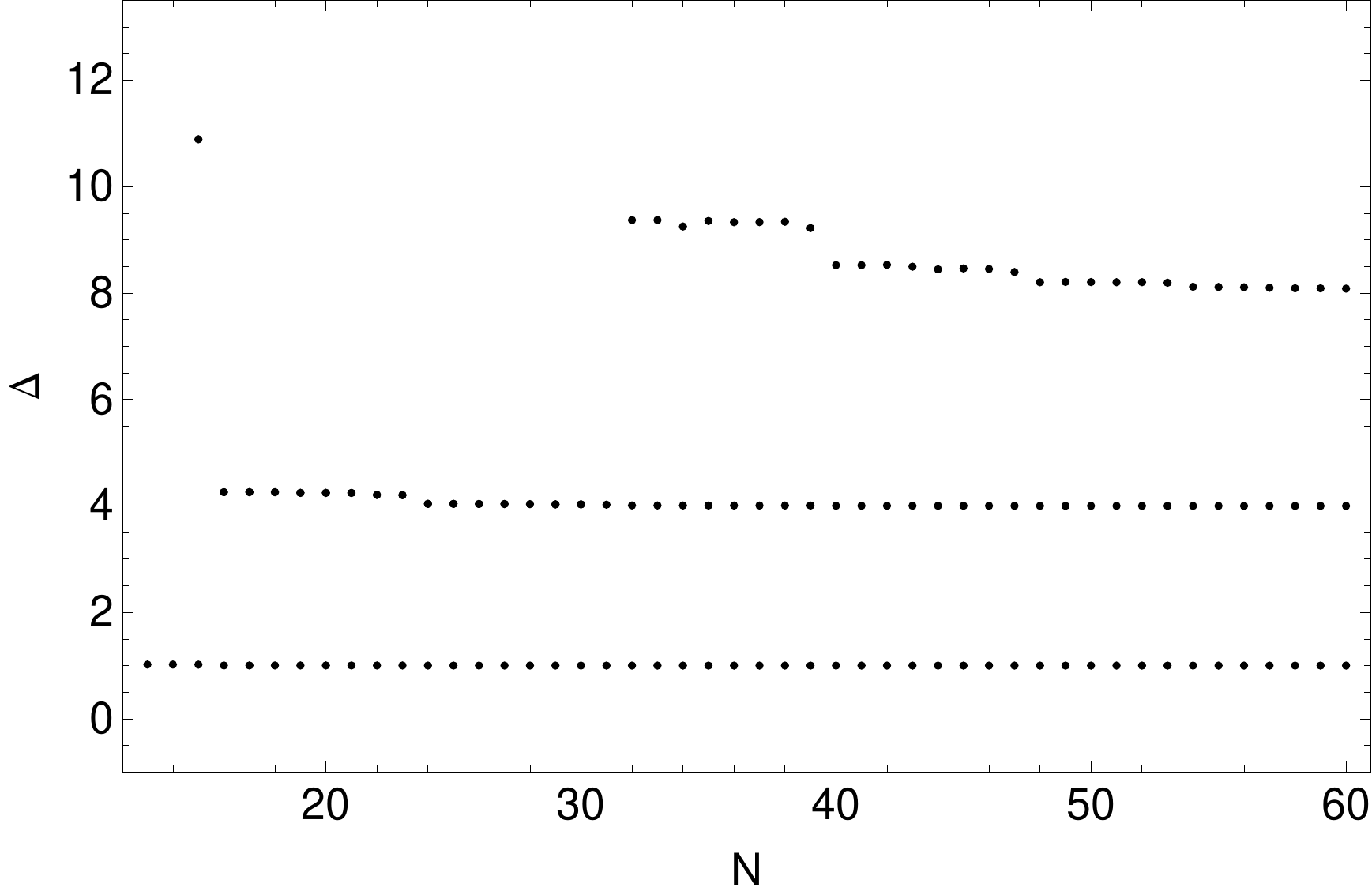}
\caption{Evolution of the zeroes of the extremal functional in the scalar channel.}

\label{fig:IsingOperatorsSpin0}
\end{figure}
It should be clear that increasing $N$ stabilizes the operator dimensions. The pattern is that as $N$ increases, the functional develops new zeroes, and shifts the positions of old ones. It is also apparent that there are discrete shifts in the spectrum at special values of $N$. These correspond precisely to points where $\Delta_\epsilon$ jumps in plot \ref{fig:epsIsing}. In any case the zeroes seem to eventually stabilize for sufficiently high $N$. This strongly suggests that increasing $N$ yields a higher and higher number of stable operators, which presumably (and our data so indicates) correspond to actual physical operators in the correlator. 

With this input we can address the issue of operator convergence. The idea is to compare how the value of the operator changes between two successive jumping points, and proclaim an operator to have ``stabilized'' only if it varies by less than some fixed amount. In this work we shall take this amount to be $1\%$, which seems to lead to reasonable results. Summarizing we propose the following working criteria:
	\begin{itemize}
	\item {\bf Criterion 1:} {\em An operator has converged, and should be trusted to appear in the actual correlator if $\delta \Delta/\Delta<1\%$. In this note $\delta \Delta\equiv \Delta_{N=60}-\Delta_{N=58}$, the positions of the two last jumping points.}
	\end{itemize}

	\subsection{Computing OPE Coefficients from the Spectrum} \label{sec:computeOPE}

The next step is to compute the OPE coefficients. For any given $N$ the extremal functional has a number of zero vectors $Z$, which generically is smaller than $N-1$. This happens not because there is some degeneracy between conformal blocks, but because many of the components of the functional are actually probing the semi-polyhedral cone along smooth directions. Such directions do not lead to any new zeroes, but rather to small improvements in the positions of those which are already present\footnote{Figure \ref{fig:2dIsing2der} shows an example of a semipolyhedron. The situation described would occur if the origin would cut across the curved scalar line (in blue). Then although there we are in 2d, there would only be one zero.}.
This means that the number of parameters is smaller than the number of constraints and so naively trying to directly solve the crossing symmetry constraints does not lead to a solution because of small numerical errors. 

To circumvent this difficulty we adopt the following procedure.  We select some subset of the total components and restrict ourselves to this lower dimensional vector space.  By restricting to only these components we can then find a set of OPE coefficients which is optimal, in the sense of minimizing the value of the crossing symmetry constraints:
\bea\label{OPELP}
	\mbox{OPE Coeffs}=\mbox{Min}_{\{\lambda_i\}}\left(\sum_{V_i:~ \phi\cdot V=0} \lambda_i \, V_i-\mathds 1\right).
	\eea
We minimize here over the value of the OPE coefficients, $\lambda_i$, where $i$ labels vectors which are zeroes of the extremal functional, $\phi \cdot V_i = 0$. Note, here we take {\em all} zeroes of the functional, not only those deemed to have stabilized.  Moreover, all the vectors appearing above have been projected onto a lower dimensional subspace of the original set of derivatives (so they have less than $N$ components).

The minimization problem \reef{OPELP} is also a linear programming problem which can be easily solved given a set of vectors. The solution is unfortunately sensitive to the choice of components used and, as remarked above, using too many components generically means no reasonable solution can be found.  It is clear that this procedure is not uniquely defined, and other choices are possible with varying quality of results. However it is important to emphasize that we have done no  special fine-tuning here in order to reproduce the expected results.  Rather there is an independent metric for determining the quality of a set of OPE coefficients, namely the value of the minimum above, and it is this metric we use to select how many components we keep from the original vector.  In practice we have found that good results for the OPE coefficients can be obtained by minimizing the above taking into account only the first $Z+1$ components of the zero vectors $V$. 

At the end of this procedure we have a hypothetical spectrum, comprising a set of operators - zeroes of the extremal functional - and their respective OPE coefficients. We can do this for any $N$, and can plot the evolution of the operator dimensions, as discussed above, and also the evolution of the respective OPE coefficients. We show this for a few selected operators in figure \ref{fig:OPEconv} of appendix \ref{plots}.
The plots clearly show that the OPE coefficients generically oscillate early on before stabilizing at a definite value. We can use this data to derive estimates on the errors of the OPE coefficients. We do this by seeing how much OPE coefficients vary between jumping points. If this variation is smaller than some cut-off we can include this operator in our hypothetical spectrum together with an error estimate. The motivation for doing this is that, although our method gives a large number of operators and OPE coefficients, not all of them should be trusted as they have not necessarily converged. To summarize, we have our
\begin{itemize}
	\item {\bf Criterion 2:} {\em An OPE coefficient $\lambda_{\mathcal O}$ has converged, and should be trusted to be approximately correct if $\delta \lambda_{\mathcal O}/\lambda_{\mathcal O}<15\%$. In this note $\delta \lambda_{\mathcal O}\equiv |\lambda_{\mathcal O_{N=60}}-\lambda_{\mathcal O_{N=58}}|$, the positions of the two last jumping points.}
	\end{itemize}
Notice that it only makes sense to talk about OPE convergence if there is an associated operator, and so criterion 2 can only be applied on operators satisfying criterion 1. The large disparity in our choice of cutoffs is mostly due to the fact that we find that operator dimensions converge more quickly than OPE coefficients, and so having a small cutoff for OPE coefficients would be too restrictive. Together, our operator dimension and OPE convergence criteria provide us with a simple way to determine whether convergence has occurred yet, and hence whether we should trust that a given operator and OPE coefficient are actually correct for some hypothetical, unknown CFT.

\subsection{Results}

In table \ref{tab:lowlyingspectrum} we present some of the operators and respective OPE coefficients obtained with our methods, side by side with those corresponding to the 2d Ising model. 
\begin{table}[htbp]
  \centering
  \caption{Low-lying 2d Ising spectrum}
  \ \\
      \begin{tabular}{|c|c|c|c|}
    \hline
    \multicolumn{2}{|c|}{Exact} & \multicolumn{2}{c|}{EFM} \bigstrut\\
    \hline
    ($\Delta$,L) & OPE   & ($\Delta$,L) & OPE \bigstrut\\
    \hline
    (1,0) & 0.5   & (1.0000025,0) & 0.4999997 \bigstrut\\
    \hline
    (4,0) & 0.015625 & (4.00030,0) & 0.0156241 \bigstrut\\
    \hline
    (2,2) & 0.1767767 & (2,2) & 0.1767772 \bigstrut\\
    \hline
    (6,2) & 0.00262039 & (5.99787,2) & 0.00261753 \bigstrut\\
    \hline
    (4,4) & 0.0209531 & (4,4) & 0.0209626 \bigstrut\\
    \hline
    \end{tabular}%
  \label{tab:lowlyingspectrum}%
\end{table}%
The agreement is impressive, especially for the low-lying operators. Our full results are presented in tables \ref{tab:spec1} and \ref{tab:spec2} of appendix \ref{plots}., including the estimated errors of our procedure. Our convergence criteria select about a dozen operators together with estimates on the error of their OPE coefficients, and comparison with the exact results show that indeed these operators are correct to very good precision, typically less than $1\%$. What this means is that for future applications, where the exact spectrum of the CFT under analysis is not actually known, we can use our method and convergence criteria to determine the low lying spectrum with reasonable confidence. 

To summarize our full set of results, we obtain:
\begin{itemize}
\item The correct value for the dimension and OPE coefficient of the operator $\epsilon$ to 6 digits accuracy.
\item  The correct central charge to within $0.0005\%$.
\item 7, 10, 15 and 19 operators with OPE coefficients correct within $0.01\%, 0.1\%, 1\%$ and $10\%$ respectively. Our highest operator has dimension 15, spin 14 with $30\%$ error on the OPE coefficient.
\item Our convergence criteria work especially well: they select out a subset of 19 operators whose OPE coefficients have converged; of those 18 reproduce the correct OPE value to within $10\%$. So, had we not known the exact spectrum our criteria would've nevertheless picked 18 operators which were actually correct to less than $10\%$. 
\item Our OPE variation error estimates are in general comparable in magnitude, or larger, than the actual errors. 
\end{itemize}
\section{Discussion}

In this note it was our goal to show that the extremal functional method is a numerically efficient way of determining the spectrum of hypothetical CFTs lying on the boundary of parameter space. We decided to focus our efforts on the two-dimensional Ising model, and in particular on an analysis of the particular correlator $\langle \sigma \sigma \sigma \sigma\rangle$. We find, with minimal computational power, more than a dozen operators and OPE coefficients to better than $1\%$ accuracy, and are quite confident this result can be extended to several dozens. Quite remarkably, we have in a way rediscovered the Virasoro symmetry in 2d by imposing only global conformal symmetry, as shown by integer spacing in the spectrum that we compute. 
These results represent a first step towards a more general method for determining the spectrum of CFTs, possibly given some minimum amount of data. A remarkable fact is that EFM explicitly shows that crossing symmetry, based on a single data point, namely the dimension of $\sigma$, allows us to reconstruct an in principle infinite\footnote{In practice as many as your computer or cluster can handle.} number of operators {\em and} their OPE coefficients, in one fell swoop. That this data is encoded in a single correlator is perhaps quite unexpected but our work indeed suggests it to be the case.

\subsection{Practical Limitations}

As we are advocating the EFM as a practical numerical approach to solving {\em unknown} CFTs such as the 3d Ising model it is important to understand the practical limitations of the method and this was in large part a motivation for this work.

It is first important to emphasize that determining the spectrum from the boundary point and determining the OPE given a particular spectrum are in fact two independent problems.  From the structure of the plots in table \ref{fig:IsingOperators} it is clear that there is a relatively unique criterion for convergence of operators and the only ambiguity lies in e.g. setting a cutoff in the variation of the operators as $N$ increases.  As mentioned above in this work we found that operators that varied by less than 1\% across jumps seemed to have stabilized and correspond to expected operators in the 2d Ising spectrum.  

The procedure for determining the OPE coefficients, given a putative spectrum, is subject to an ambiguity.  As explained above this is because the cone is semi-polyhedral rather than polyhedral so adding a new dimension (component to the vector) does not necessarily generate a new vertex on a face.  Rather the cone can curve continuously along such a direction.  As this generically yields less zeroes, $Z$, than constraints, $N-1$, the crossing symmetry constraint problem is over-determined (albeit by small corrections).  We find that using the first $Z+1$ components of the functional, ordered by the value of $n+m$ (the total number of transverse and longitudinal derivatives) yields very good results. Our choice is conservative, in the sense that it is possible that choosing a different subset of components can lead to further improvements. Notice that ``improvements'' here refers to how well our solution satisfies the full set of $N-1$ constraints, and so this is an independent criteria requiring no fine-tuning.  That is, the degree to which a set of computed OPE coefficients satisfy crossing symmetry is something which can be computed independently of the knowledge of the exact results for these coefficients, and so we can use it as a guide when applying our methods to unknown CFTs.

A different set of issues arise when pushing our methods to higher and higher orders. One is that higher derivatives of the conformal blocks vary dramatically depending on the conformal dimension. This can lead to numerical problems, since we are comparing numbers across several orders of magnitude. Due to the scale of the problem and the nature of the software used numerical precision is inherently limited (to a total range of approximately 16 orders of magnitude) and this introduces significant numerical issues when going to high derivative order (which is necessary to extract more of the spectrum).  There are many ways to potentially alleviate this problem  such as using lower-order derivative expansions around several separated points or changing to another basis altogether.
A related issue is the question of how precisely we can determine the dimensions of the operators.  As discussed in \cite{Pappadopulo:2012jk} there is an exponential fall-off in the contribution of higher dimension operators to a given correlator and this translates into exponentially suppressed dependence on the precise operator dimension.  Including a huge number of very thinly spaced operators at high dimension leads to a large degeneracy in the linear program as these operators are hard to distinguish and this results in  additional numerical instabilities. 

Finally let us consider what the general implications of the estimates in \cite{Pappadopulo:2012jk} are for our approach. There it was shown that almost everywhere in the $u,v$ plane the convergence of the conformal block expansion is exponentially fast. In light of these results, the low lying operators in a correlator approximate it extremely well everywhere, but not in all channels simultaneously: close to the boundary points $u=1, v=0$ and $u=0, v=1$ convergence becomes  slow for one of the channels. However, this is not an issue for us, since our method obtains data locally around $u=v=1/4$, and given a solution to crossing we can sum up conformal blocks in the channel most convenient to us.
In summary, for all practical purposes the current accuracy of our methods  should be sufficient for determining the full correlator in a large region to extremely high precision. On the other hand, this result also implies that since OPE coefficients decrease exponentially fast at large conformal dimension, it will be extremely hard in practice to pin them down to high precision.

\subsection{Theoretical Limitations}

There are two obvious limitations of our method for determining the spectrum of a CFT. The first is that we are restricted to exploring CFTs which sit at the boundary of some exclusion plot in parameter space. It therefore seems that we require quite a bit of luck in that we have to hope that the CFTs of interest do not lie down in the interior of the allowed region. The second is that we need to have some information about the CFT, such as the dimension of a scalar operator, which tells us where we are located along the boundary curve. Let us address these two problems in turn.

Firstly, we notice that the boundary of an exclusion plot can be modified by imposing additional conditions, as was shown in \cite{ElShowk:2012ht}. What this means in practice is that if we have some additional data about the theory under study (such as the dimensions of a few extra operators), we can impose these as extra constraints to obtain a new boundary. The hope then is that a suitably picked constraint can place the desired CFT on the boundary, at which point we can apply our methods without restrictions. A different way of constraining the boundary might be to consider a larger subset of correlation functions, as we discuss further below. 

As for the second problem (which is not independent from the first), we need an initial, known value of $\Delta_\sigma$. The accuracy to which this number is known affects all remaining results. In the 2d Ising model we do know this number exactly, but we are not as fortunate e.g. in the 3d Ising model, where $\Delta_\sigma$ is known to about 3 digits accuracy. Our method necessarily propagates the error in the determination of this number to all other results. On the other hand, given this single piece of data our method determines a large number of other operators and their OPE coefficients. This suggests that other kinds of numerical efforts, such as lattice computations, should be focused on examining two point functions of a single scalar operator; a significantly easier problem than e.g. considering three-point functions. A last comment on this issue is that we might be extremely lucky and find that the CFT is located on a special feature in the exclusion plot, such as a kink. This seems to be the case for both the 2d and 3d Ising models \cite{Rychkov:2009ij,ElShowk:2012ht}. In this case, a detailed examination of such a feature can in principle lead to a high precision determination of $\Delta_\sigma$.

An important point is that the EFM only computes the solution to crossing symmetry of a single correlator. This does not allow us to claim that we have constructed a consistent CFT, since these should satisfy an infinite set of additional constraints, including crossing symmetry relations on all other correlators (an infinite set!) and global properties like modular invariance in two-dimensions. An important step in understanding how these constraints cross-pollinate would be to study the $\sigma,\epsilon$ subsector of correlation functions to check whether stronger constraints arise. Evidence that this is indeed the case is evident in the works \cite{Poland:2010wg,Vichi:2011ux,Poland:2011ey}. In particular semi-definite programming methods were used to study several correlators simultaneously to great effect. We leave this important avenue of research for future work.

\subsection{Future Directions}

Although we have chosen to focus on a single point on the boundary of parameter space, nothing prevents us from computing the spectra at other points along the curve, although whether such points can correspond to actual physical CFT's remains unclear. The reason of course is that one should demand not only crossing symmetry constraints on a single correlator, but on an infinite set of them. Nevertheless, probing the region around a given CFT can provide valuable information on the structure of its spectrum. For instance, it is well known that minimal models are very special points, as they have no negative norm states, and an infinite set of null states. They are small, isolated islands in an ocean of non-unitary theories. As such it is likely that a careful study of the spectrum generated by crossing-symmetry in the neighborhood of the minimal model points could provide signatures of their special status.

There are many interesting directions for future research. Here is a partial list:

\begin{itemize}

\item It would be nice to carefully study the spectrum and their variations close to the Ising point. Preliminary results seem to show interesting structure there which could explain the origin of the kink, which we hope to address soon. 

\item In \cite{ElShowk:2012ht} evidence was found for the existence of such a kink in dimension 3 at the position of the 3d Ising model. For this model we have limited information on the spectrum of the theory, including $\Delta_{\sigma}$ which is known only up to 3 digits accuracy. It is our hope that by studying how the spectrum changes as $\Delta_\sigma$ is varied we might be able to do better than this. Of course, once this piece of data is given, we hope to reconstruct the low lying spectrum of the 3d Ising model to better accuracy than what is currently known in the literature.

\item Our method works for any value of dimension, even fractional ones. As such it would be interesting to follow the flow of the spectrum from the 4d free theory all the way to the 2d Ising point along the Wilson-Fischer line. A first, less ambitious step would be to reproduce the results of the epsilon-expansion \cite{Wilson:1971dc}.

\item We need to explore how to best impose extra cuts in the spectrum to isolate CFTs of interest. A possible test case would be to consider the tricritical Ising model, which has an operator $\sigma$ with dimension $3/40$. In the $\sigma \sigma$ OPE the lowest scalar appearing, $\epsilon$, has dimension $1/5$. This pair of points is indeed below the 2d bound of figure \ref{fig:2dScan}.

\item It would be interesting to determine whether exploring a larger subsector of correlators can lead to stronger exclusion plots. Technically this requires major developments, as we must now use semi-definite programming methods \cite{Poland:2011ey}. On the other hand such methods do away with having to discretize the space of conformal blocks, leading to cleaner results for the spectrum.

\item Our methods are directly applicable to superconformal theories. In particular we can readily determine the spectra of theories located on the boundaries of the exclusion plots found in \cite{Poland:2010wg}.

\item There is a lot of room for technical improvements but also theoretical ones. Is there a better basis for the functional other than derivatives? Can we have a better theoretical understanding of the extremal functional method by analysing the structure of conformal blocks?

\end{itemize}

\acknowledgments{We would like to thank many useful discussions with  P. Liendo, L. Rastelli, B. van Rees, and especially D. Poland, D. Simmons-Duffin, S. Rychkov, and A. Vichi. We have also benefitted from a stimulating environment at the Back to the Bootstrap 2 conference, and thank Perimeter Institute and the organizers for their hospitality. MP acknowledges funding from the LPTHE at Univ. Pierre et Marie Curie, Paris, and from D.O.E. grant DE-FG02-91ER40688.  The work of S.E. is supported primarily by the Netherlands Organization for Scientific Research (NWO) under a Rubicon grant and also partially by the ERC Starting Independent Researcher Grant 240210 - String-QCD-BH.}

\appendix

\section{Implementation of linear programming}
\label{lp}
The crossing symmetry constraints \reef{cross2} can be written as
\bea
\sum_{\Delta,l} \lambda_{\mathcal O_{\Delta,L}}^2\left(v^d G_{\Delta,l}(u,v)-u^d G_{\Delta,l}(v,u)\right)=u^d-v^d \label{constraints}
\eea
The notation is schematic: we should really allow for an uncountable set of operators in the sum above, since the conformal dimension is a continuous parameter. In practice, however, we need to discretize $\Delta$, as will be discussed below. As mentioned previously we also discretize the functional dependence by working with coefficients of a power series: we define $u=z\bar z, v=(1-z)(1-\bar z)$ and then expand around $z=\bar z=1/2$.  More concretely we take $z=\frac 12 (1+a+\sqrt{b}), \bar z=\frac 12(1+a-\sqrt{b})$ and expand in $a$ and $b$. We use $m$ and $n$ to denote the maximum number of $a$ and $b$ derivatives, respectively. In this work we consider $m=n=7$. 

The next problem is evaluating the derivatives of the conformal blocks. In $D=2$ this is not a problem since the conformal blocks have been known for some time (see e.g. \cite{DO1}). However we want to keep our method valid for general dimension, and hence we use the algorithms developed in \cite{ElShowk:2012ht}. In practice, these algorithms require computing higher order derivatives along the $b$ direction in order to compute derivatives along $a$. We will use these extraneous derivatives in our computations, and so effectively our $m$ is actually larger than $7$.  More precisely, for a given $n$ we have a maximum number of $m$ derivatives available given by:
	\bea
	m(n)=7+2\times(7-n)
	\eea
Further, only odd numbers $m$ derivatives appear, since even ones turn out be zero. Overall the total number of derivative is
	\bea
	N_{\mbox{\tiny Max}}=\frac {(1+n_{\mbox{\tiny Max}})(1+m_{\mbox{\tiny Max}}+n_{\mbox{\tiny Max}})}2
	\eea
with $n_{\mbox{\tiny Max}}=m{\mbox{\tiny Max}}=7$, which gives $N_{\mbox{\tiny Max}}=60$. When producing results with smaller $N$, such as when we consider the evolution of the spectrum with the number of components, we order the full list of 60 derivatives by the value of $m+n$ (e.g. $(3,1)<(5,0)$), and take the first $N$ of them.


The second problem is discretizing the allowed space of conformal blocks; turning $\Delta$ from a continuous to a discrete label. We do this by computing a table of operators with dimensions up to $\Delta_{\mbox{\tiny Max}}=100$ and spins up to $50$, with a spacing of $0.1$ between dimensions. We then interpolate this table so that we can obtain vectors at intermediate dimensions. We have checked that the interpolated results approximate the exact results extremely well. 


At this point we have a large but finite set of vectors with a number of components given by $N$. The crossing symmetry constraints \reef{constraints} can now be written schematically as
	\bea
	\sum_{\Delta,L} \lambda_{\Delta,L}^2 V_{\Delta,L}=\mathds 1
	\eea
where $\mathds 1$ stands for the identity vector derived from $u^d-v^d$. To rule out solutions of the above equation we look for a linear functional which is positive on all vectors except the identity, where it should be negative. Geometrically, we are looking for a hyperplane separating out the identity vector from the cone generated by the vectors $V$. If this is possible, clearly there is no solution to \reef{constraints}. As in the work \cite{ElShowk:2012ht} we use IBM's {\tt ILOG CPLEX} optimizer\footnote{http://www-01.ibm.com/software/integration/optimization/cplex-optimizer/} to construct such linear functionals. We setup our calculations using {\tt Mathematica} and interface with {\tt CPLEX} via {\tt MathLink}.

As described in the main text, to find a bound we impose gaps in the spectrum of the putative CFT and look for the existence (or not) of a linear functional satisfying all the constraints. We use a simple bisection algorithm to determine the maximal possible value of $\Delta_\epsilon$ at $\Delta_\sigma=0.125$ and this yields a nearly extremal functional and a value for $\Delta_\epsilon^* = 1.000003$. 

In practice it is not efficient to send the entire vector table into the linear programming algorithm. Instead, we start off with a smaller table, more densely constructed at lower dimensions and successively rarefied at higher dimensions and spin.  The sparseness of our vector tables at higher dimension is justified as the functions $F_{\Delta, l}^\sigma$ vary much more quickly as a function of $\Delta$ for $\Delta$ close to $\Delta_\sigma$ (this is easily checked numerically).  

We execute our bisection algorithm and find the extremal functional. From this functional we read off the zeroes, which are our putative operators. At this point we refine our vector table by including more vectors around the positions of the zeroes. In this way we can obtain very accurate results on the position of the zeroes without using large tables. This is only possible because of our use of interpolation which allows us to obtain vectors at arbitrary dimensions with good precision. For more details on the exact choice of vectors and the resulting refined tables at each iteration the reader is referred to the notebook attached to the online version of this note.

At the end of this procedure we have a table of operator dimensions, from which we need to obtain OPE coefficients. To do this we take the corresponding vectors and look at a subset $\hat N$ of their components, the size of which we correlate with the total number of zeroes $Z$. We take $\hat N=Z+1$. Given this set of vectors we are interested in finding the set of OPE coefficients which give a best fit to equation \reef{cross2}. This can be easily done again through the use of a linear program, since, as described below eqn \reef{OPELP}, it is the solution to
	\bea
	\mbox{OPE Coeffs}=\mbox{Min}_{\{\lambda_i\}}\left(\sum_{V:~ \phi\cdot V=0} \lambda\cdot V-\mathds 1\right).
	\eea
which finally gives us the desired coefficients.

\newpage

\section{Plots and Tables}
\label{plots}
\begin{figure}[ht]
\centering
\begin{tabular}{cc}
$L=0$ & $L=2$ \\
\includegraphics[scale=0.25]{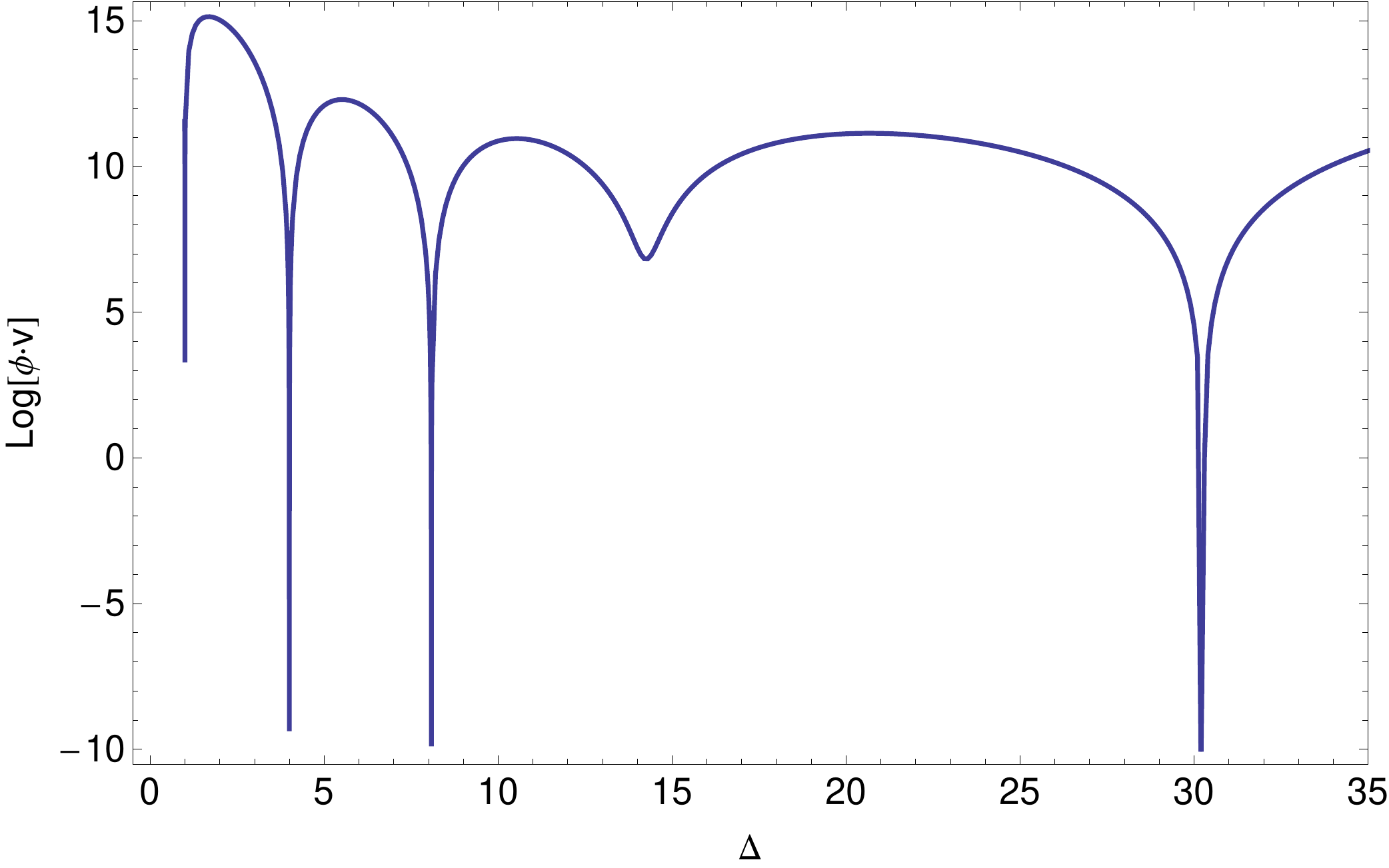}&
\includegraphics[scale=0.25]{FunctionalL2.pdf} \\
& \\
$L=4$ & $L=6$\\
\includegraphics[scale=0.25]{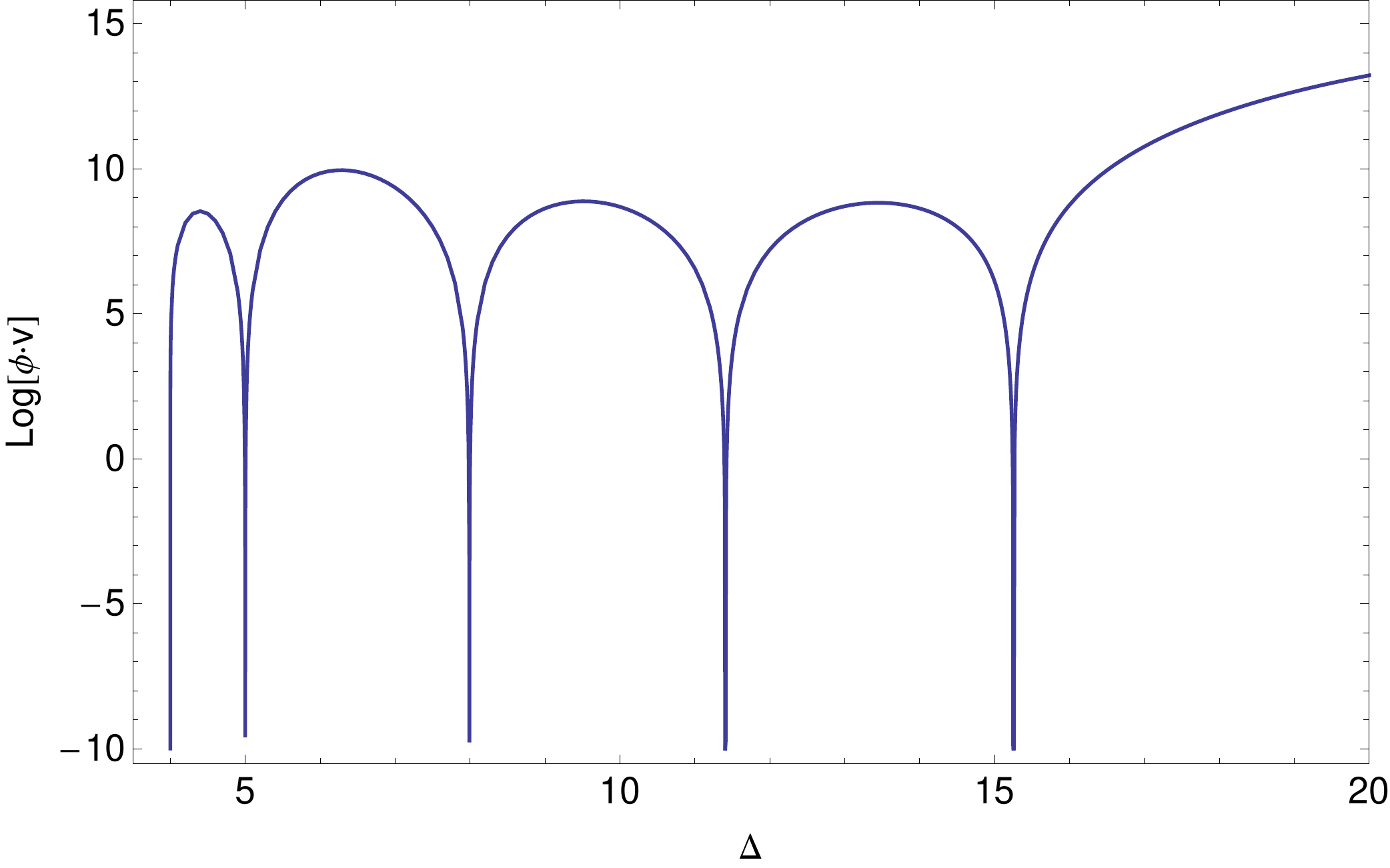}&
\includegraphics[scale=0.25]{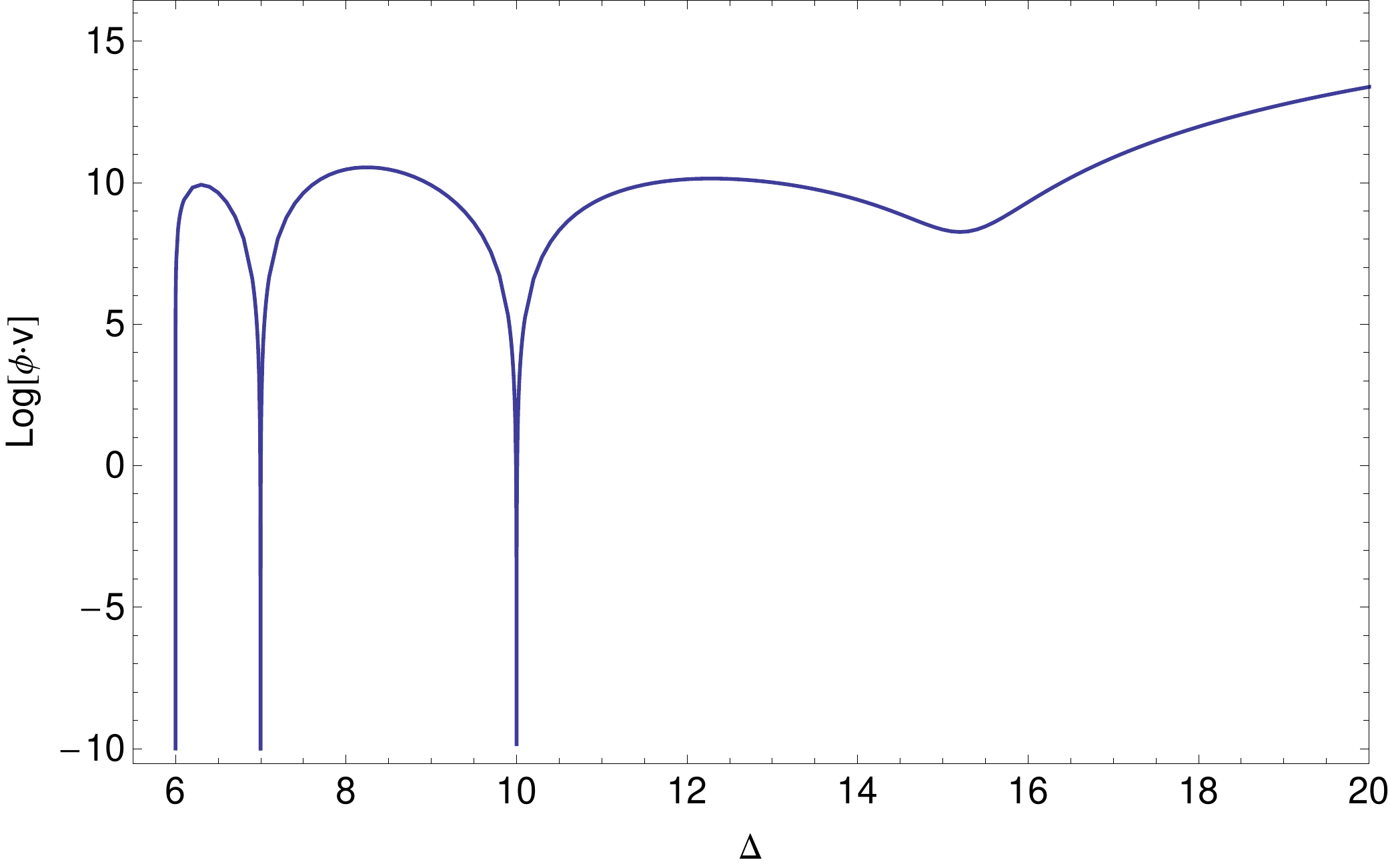} \\
& \\
$L=8$ & $L=10$\\
\includegraphics[scale=0.25]{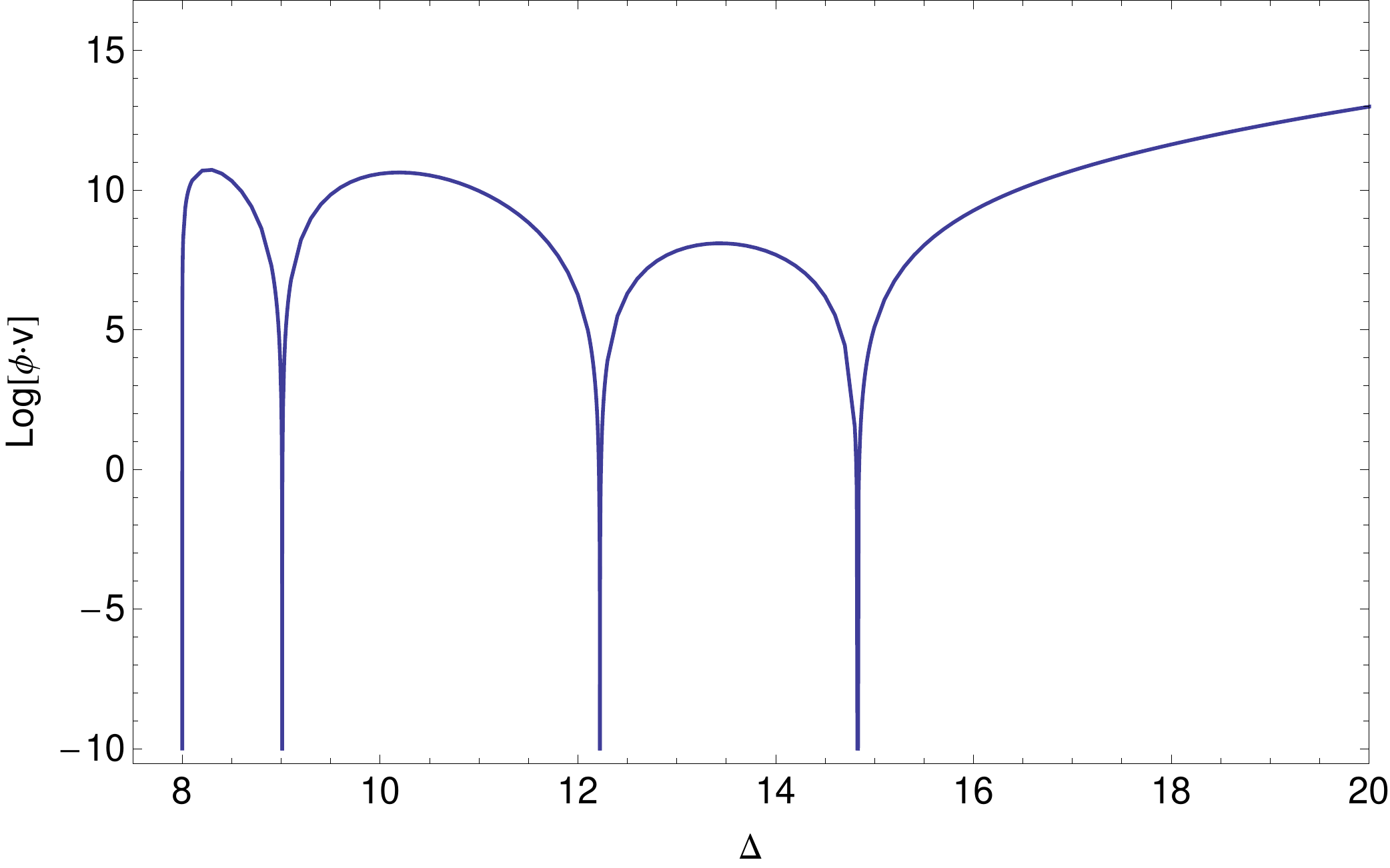}&
\includegraphics[scale=0.25]{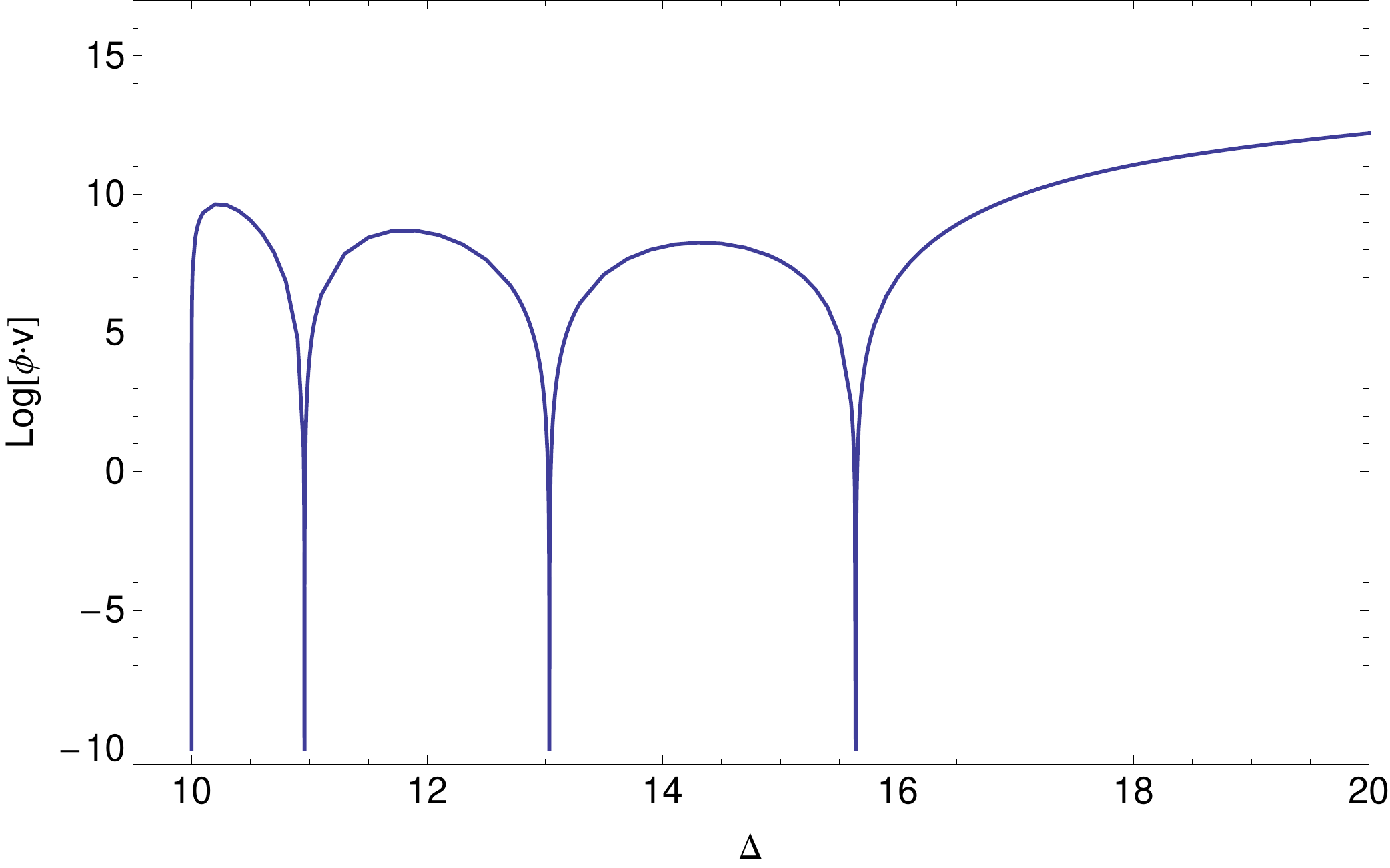} \\
$L=12$ & $L=14$\\
\includegraphics[scale=0.25]{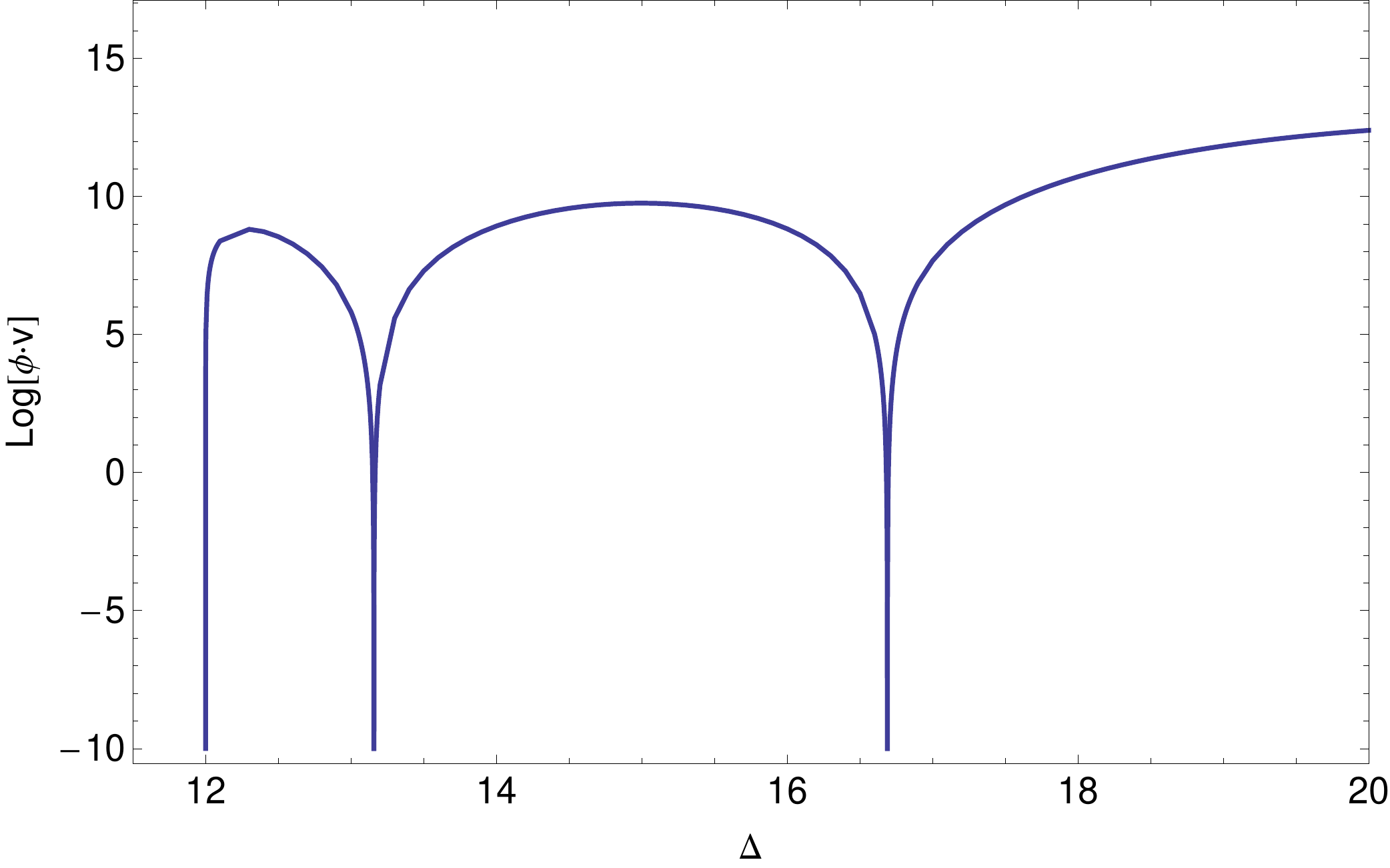}&
\includegraphics[scale=0.25]{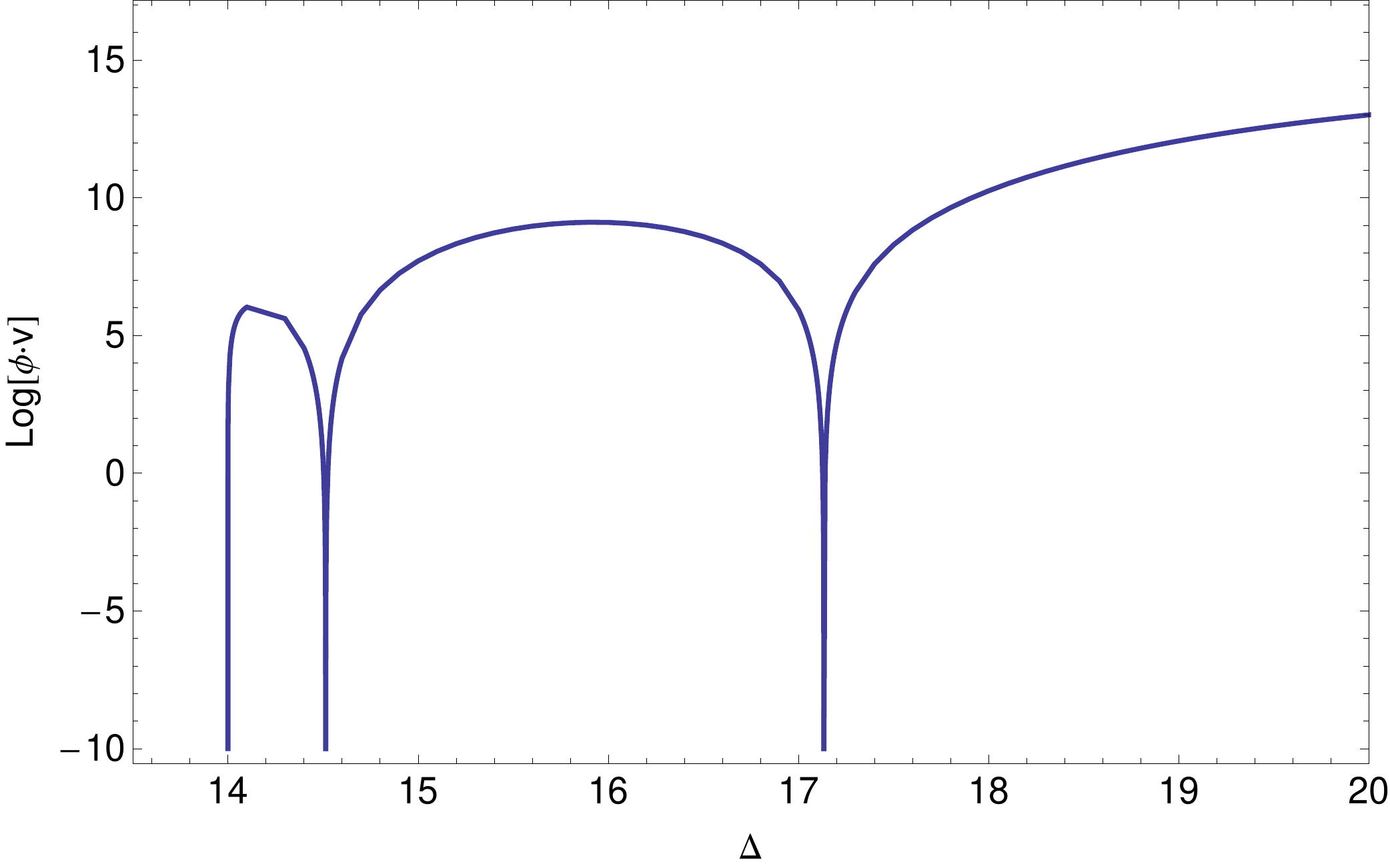}
\end{tabular}
\caption{Plots of $\phi_{\mbox{\tiny Ext}}\cdot v(\Delta,L)$. Extremal functional $\phi$ determined at the Ising model point, $d_{\sigma}=0.125$.}
\label{fig:IsingFunctionals}
\end{figure}%

\newpage

\begin{figure}[ht]
\centering
\begin{tabular}{cc}
$L=0$ & $L=2$ \\
\includegraphics[scale=0.3]{IsingL0.pdf}&
\includegraphics[scale=0.3]{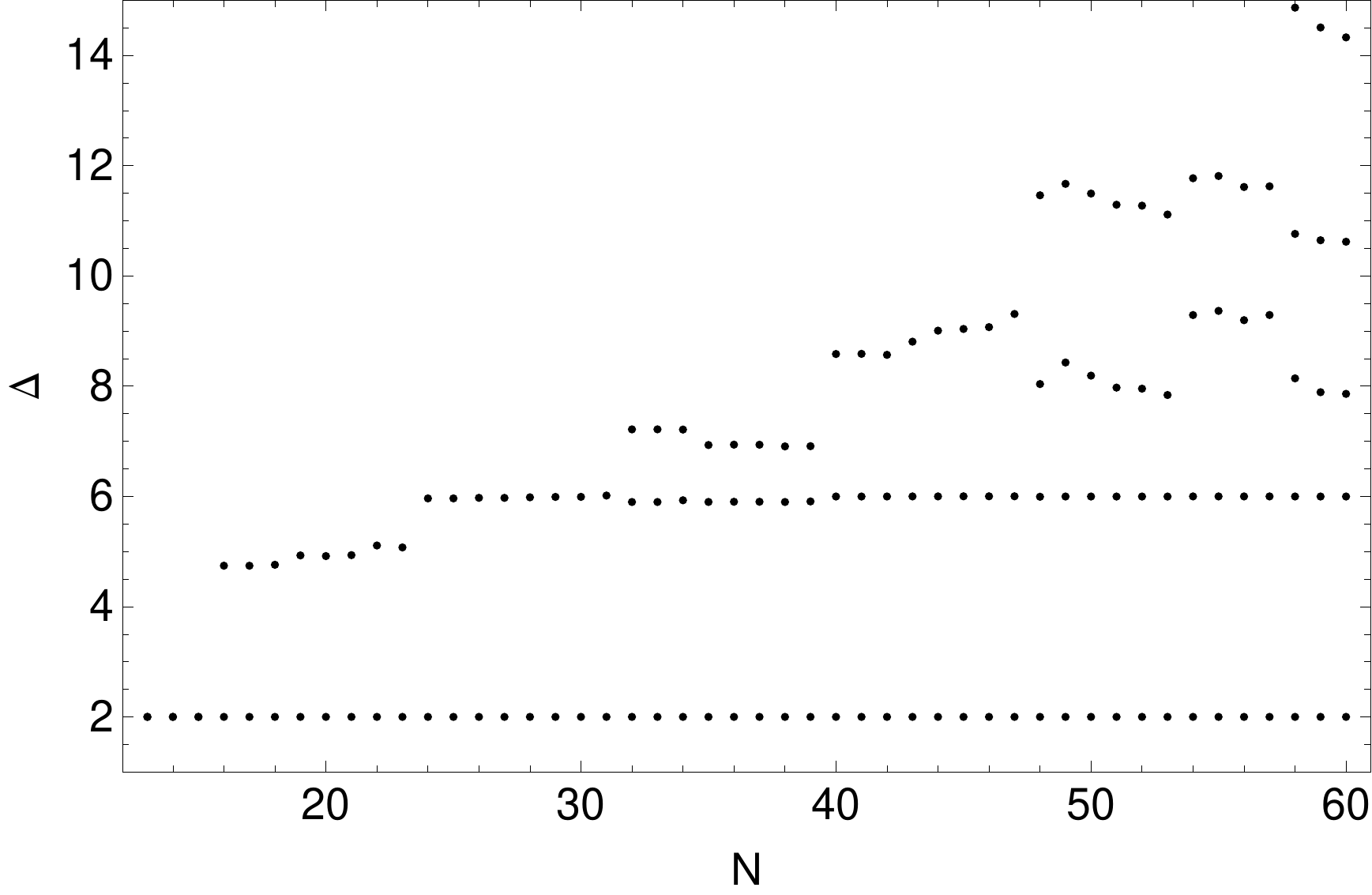} \\
& \\
$L=4$ & $L=6$\\
\includegraphics[scale=0.3]{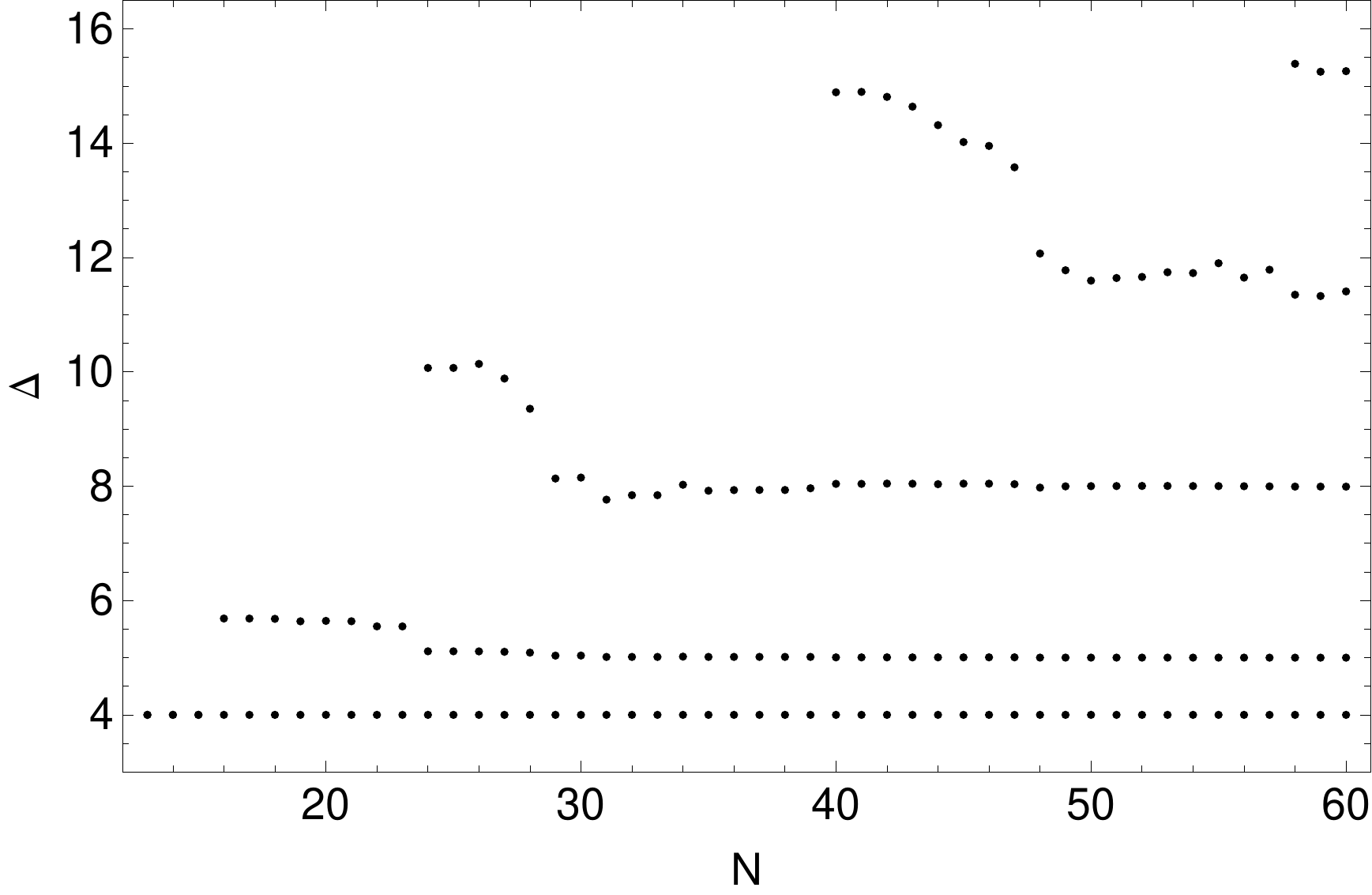}&
\includegraphics[scale=0.3]{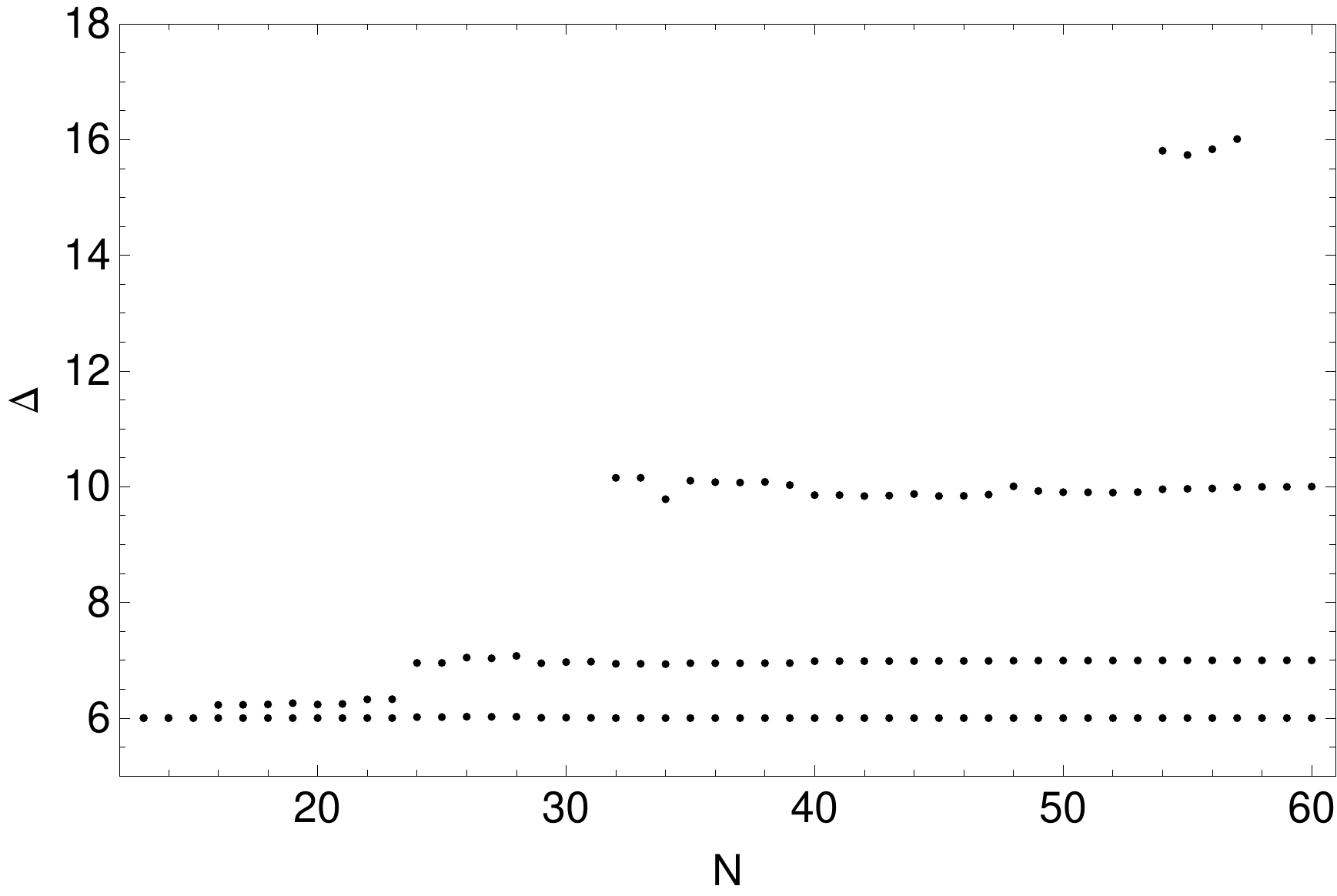} \\
& \\
$L=8$ & $L=10$\\
\includegraphics[scale=0.3]{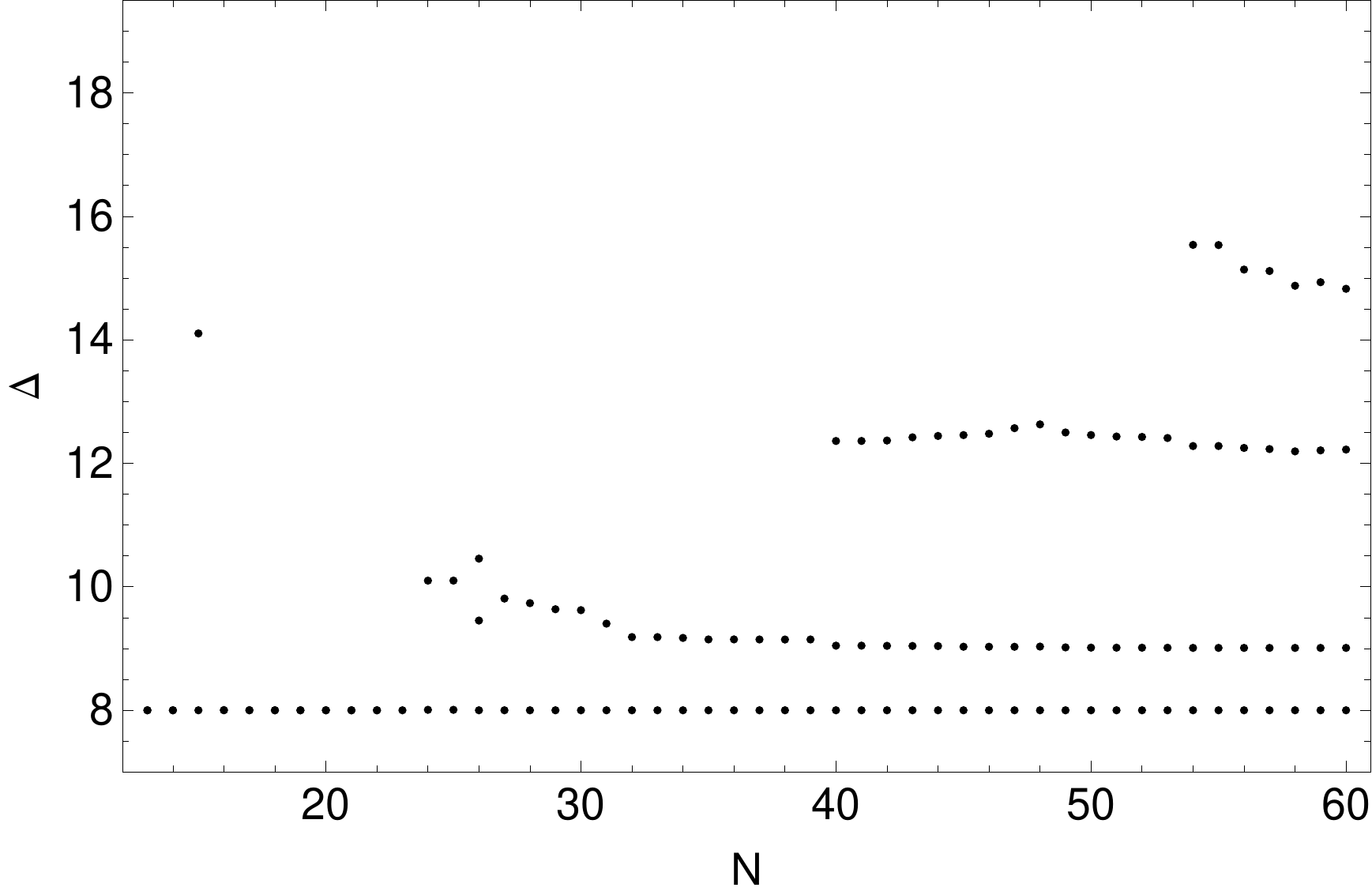}&
\includegraphics[scale=0.3]{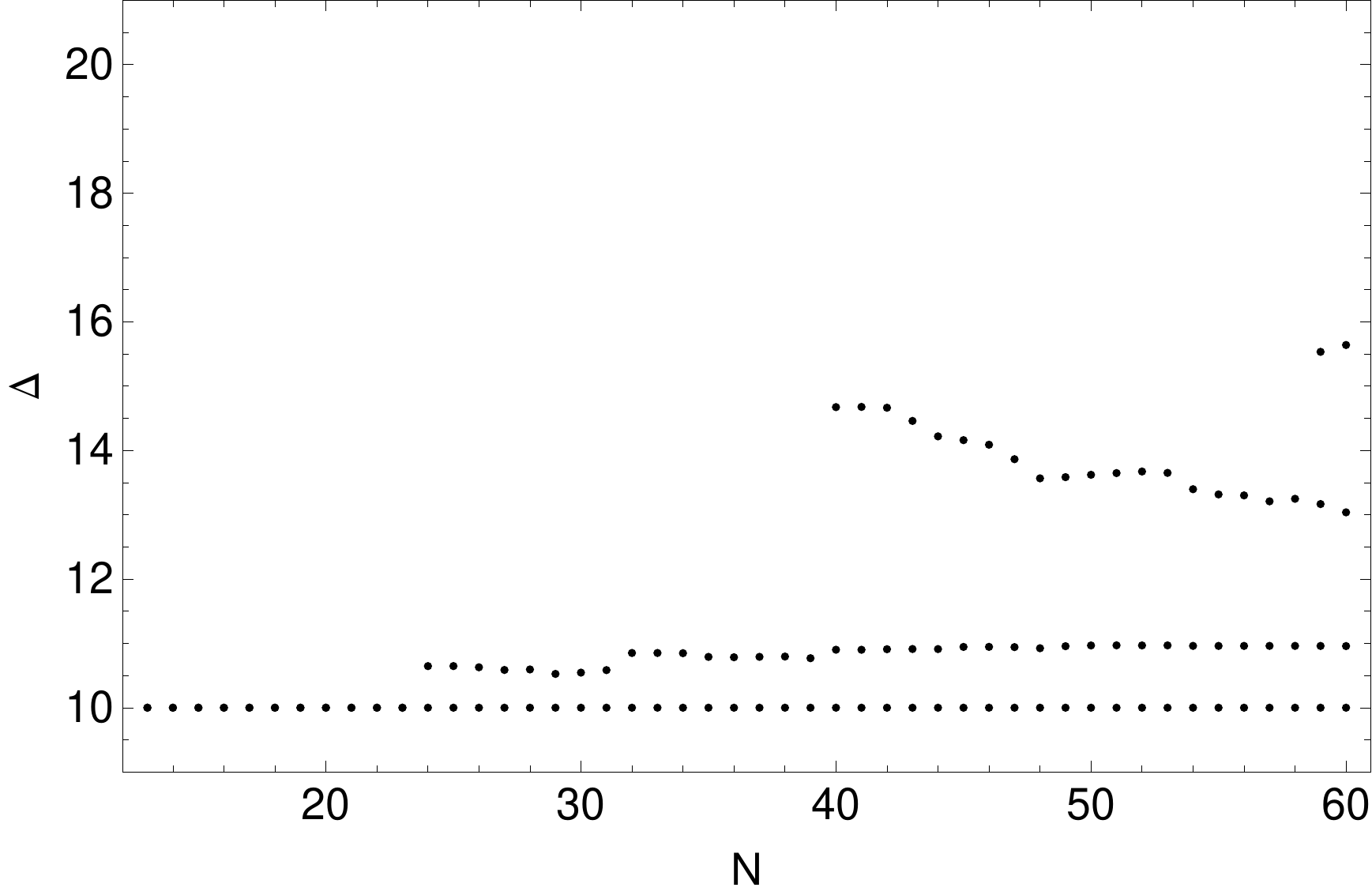}\\
$L=12$ & $L=14$\\
\includegraphics[scale=0.3]{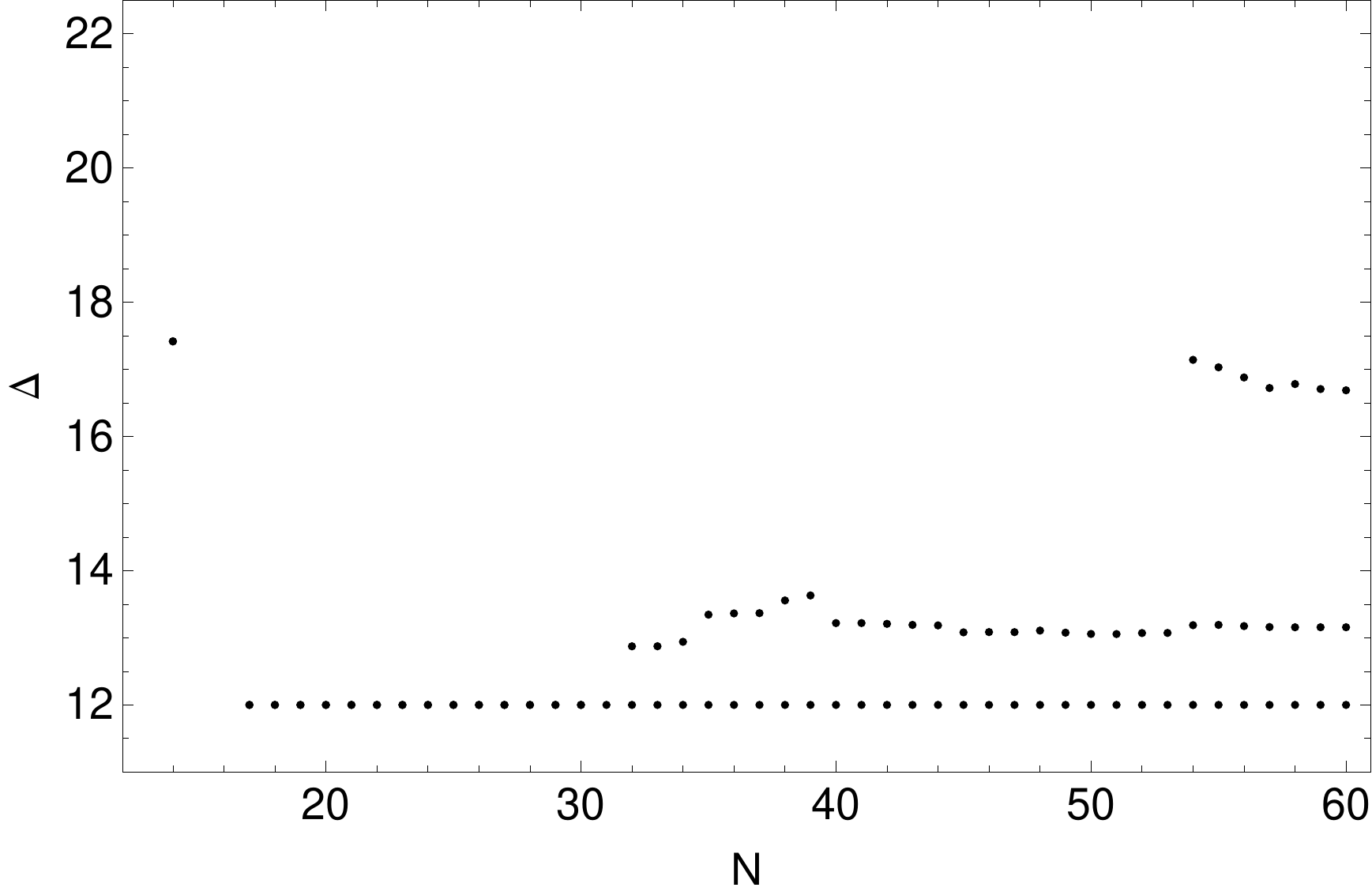}&
\includegraphics[scale=0.3]{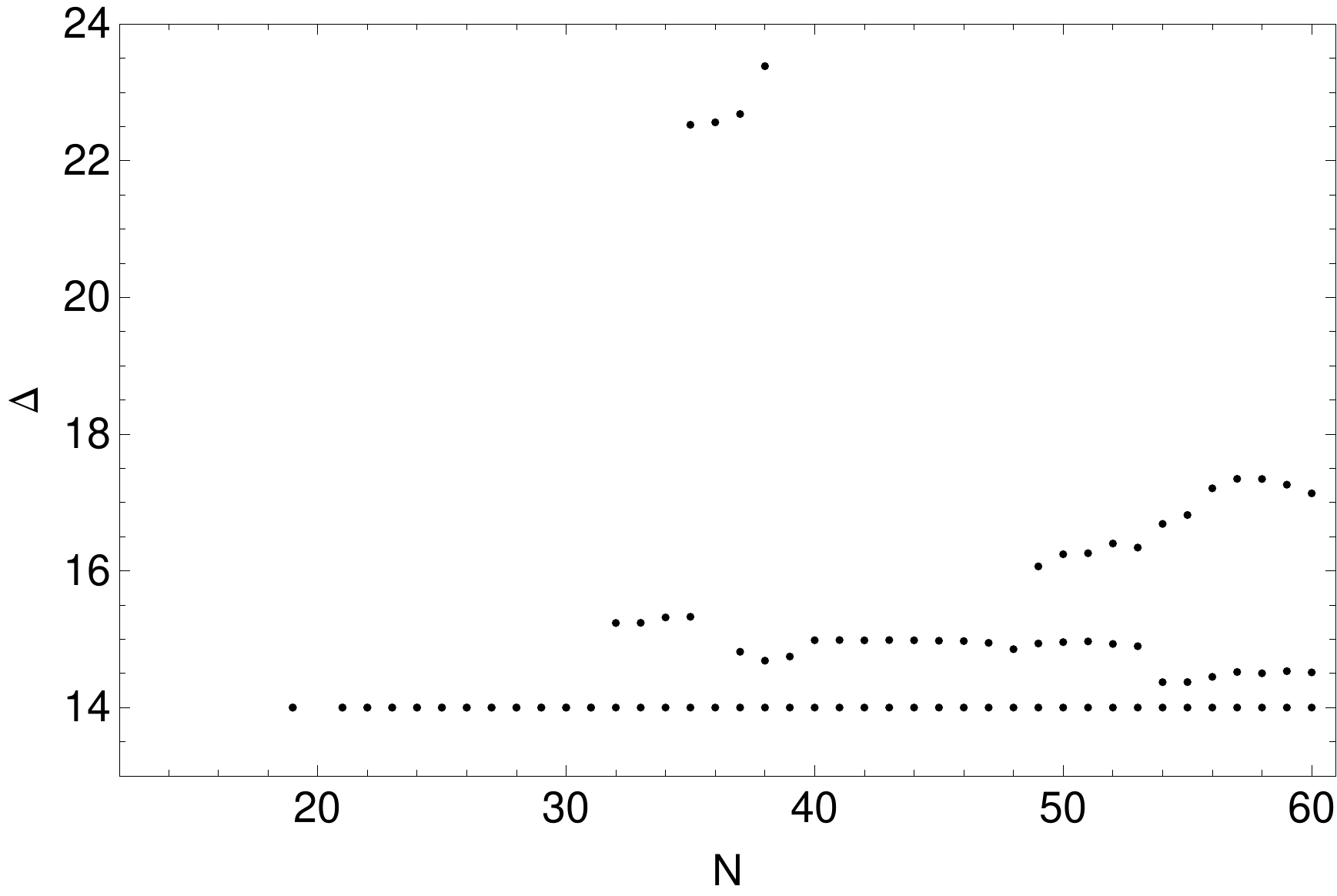}
\end{tabular}
\caption{Evolution of the zeroes of the extremal functional at the Ising point as a function of $N$, the number of components in the derivative vectors.}
\label{fig:IsingOperators}
\end{figure}

\newpage


\begin{figure}[ht]
\centering
\begin{tabular}{cc}
$(\Delta,L)=(4,0)$ & $(\Delta,L)=(8,0)$ \\
\includegraphics[scale=0.5]{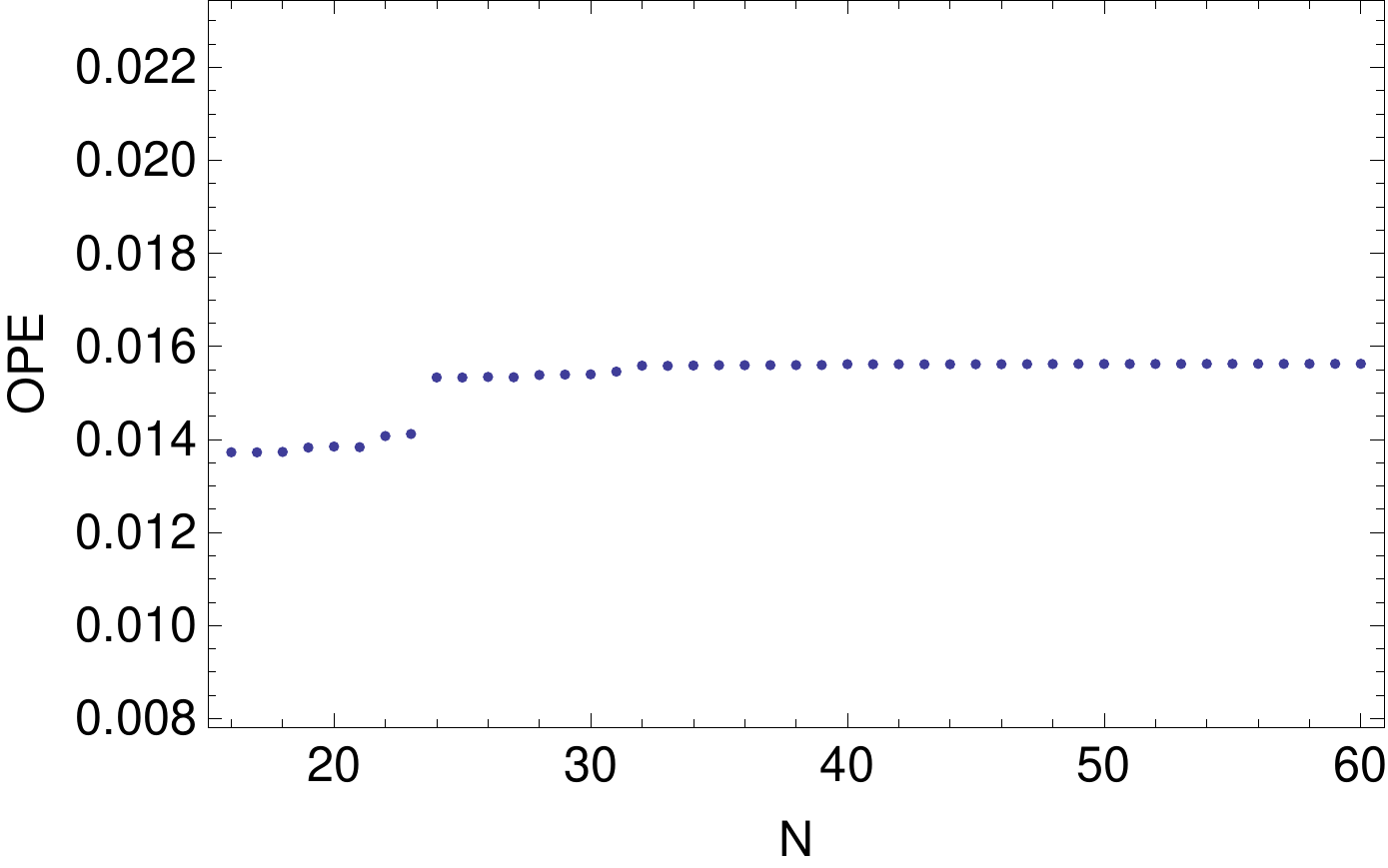}&
\includegraphics[scale=0.5]{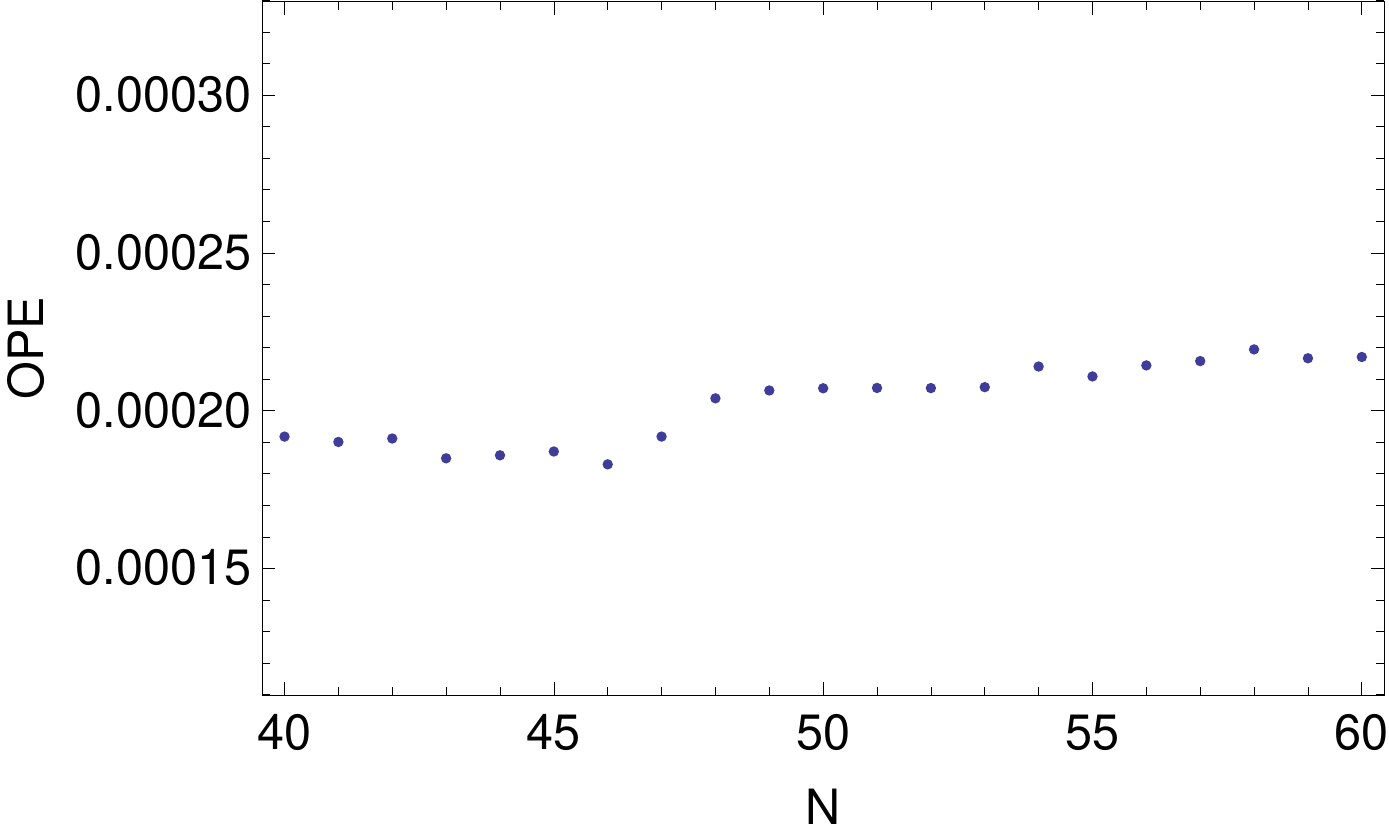} \\
& \\
$(\Delta,L)=(2,2)$ & $(\Delta,L)=(8,4)$\\
\includegraphics[scale=0.5]{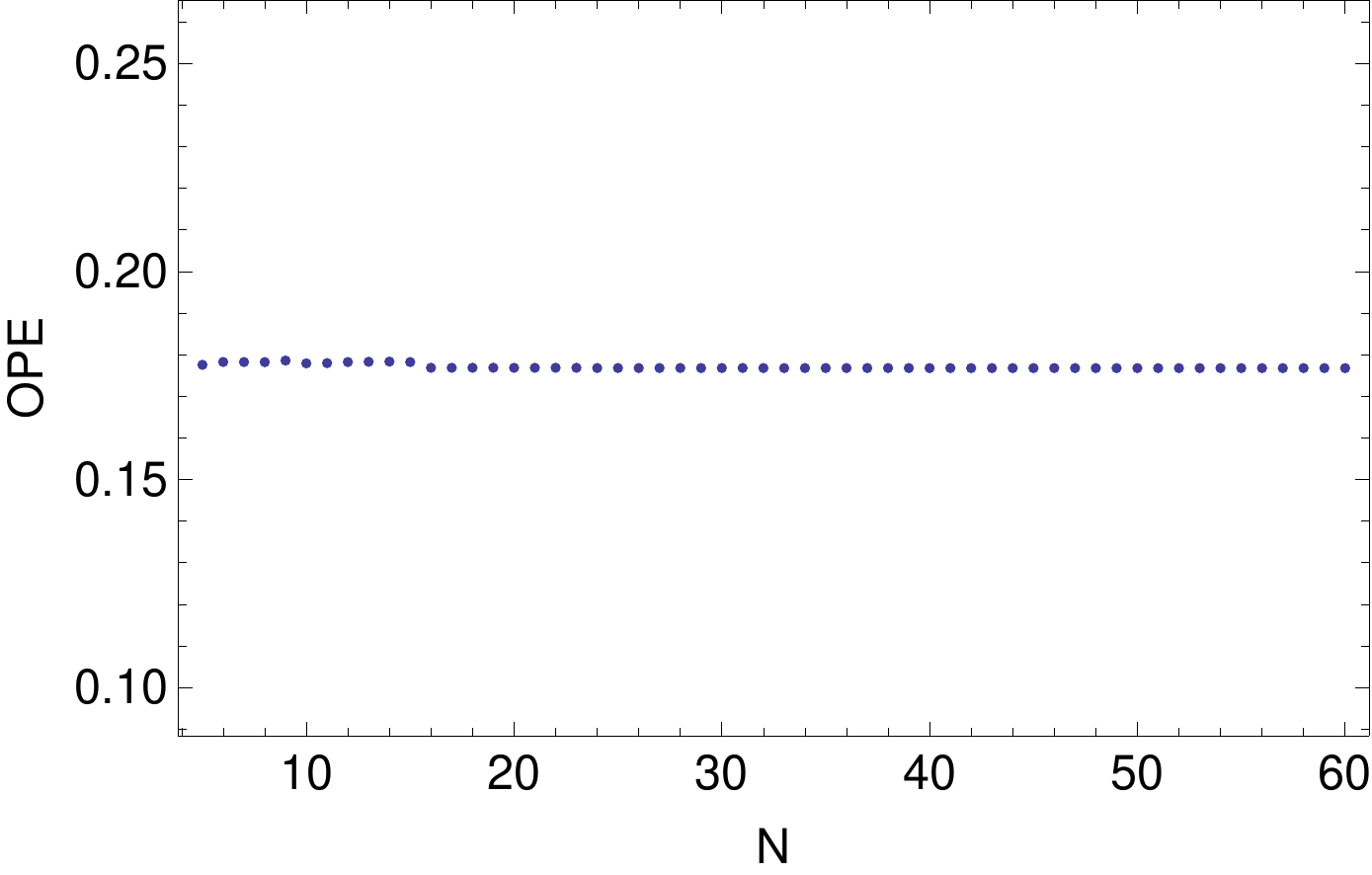}&
\includegraphics[scale=0.5]{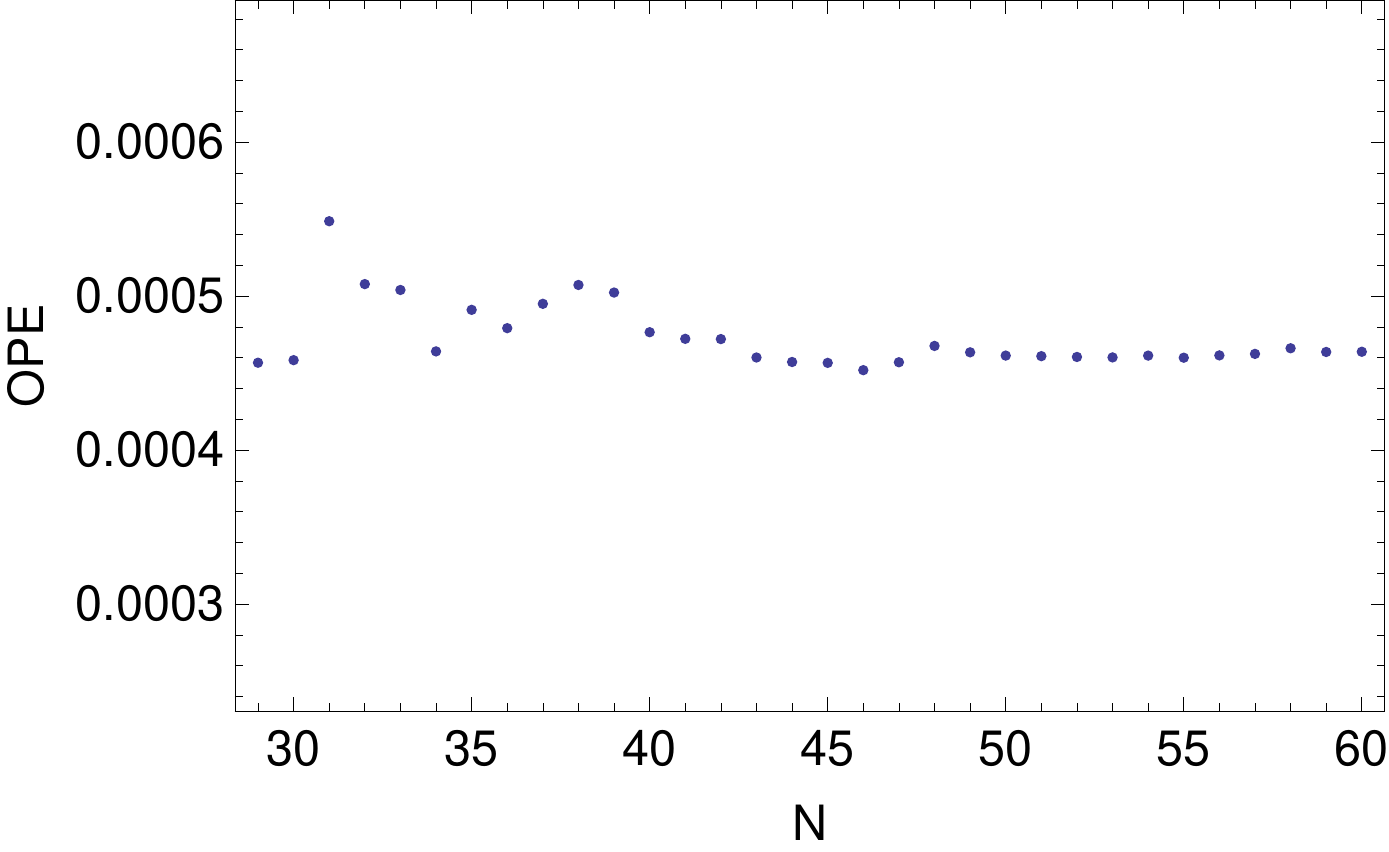} \\
& \\
$(\Delta,L)=(6,6)$ & $(\Delta,L)=(8,8)$\\
\includegraphics[scale=0.5]{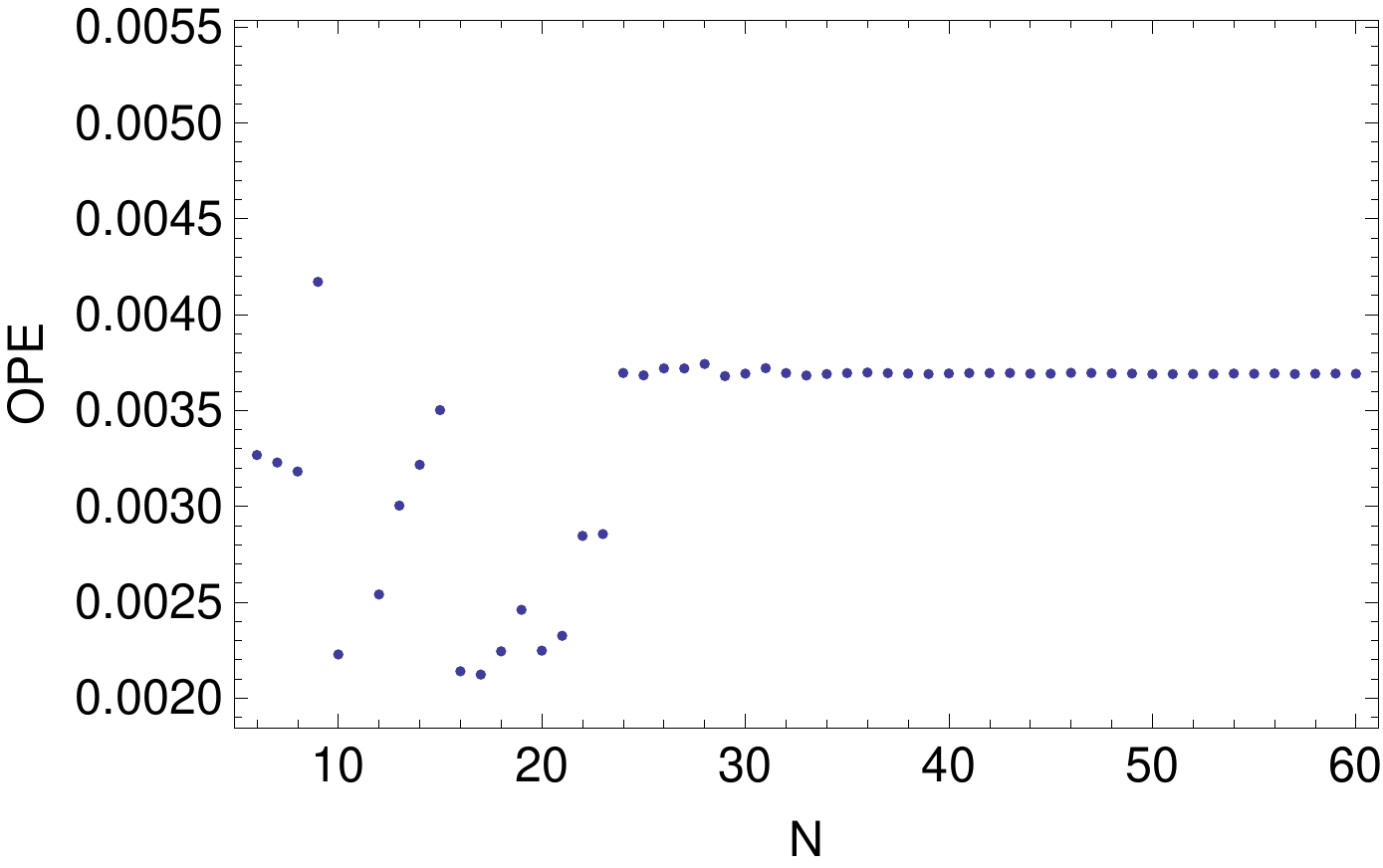}&
\includegraphics[scale=0.5]{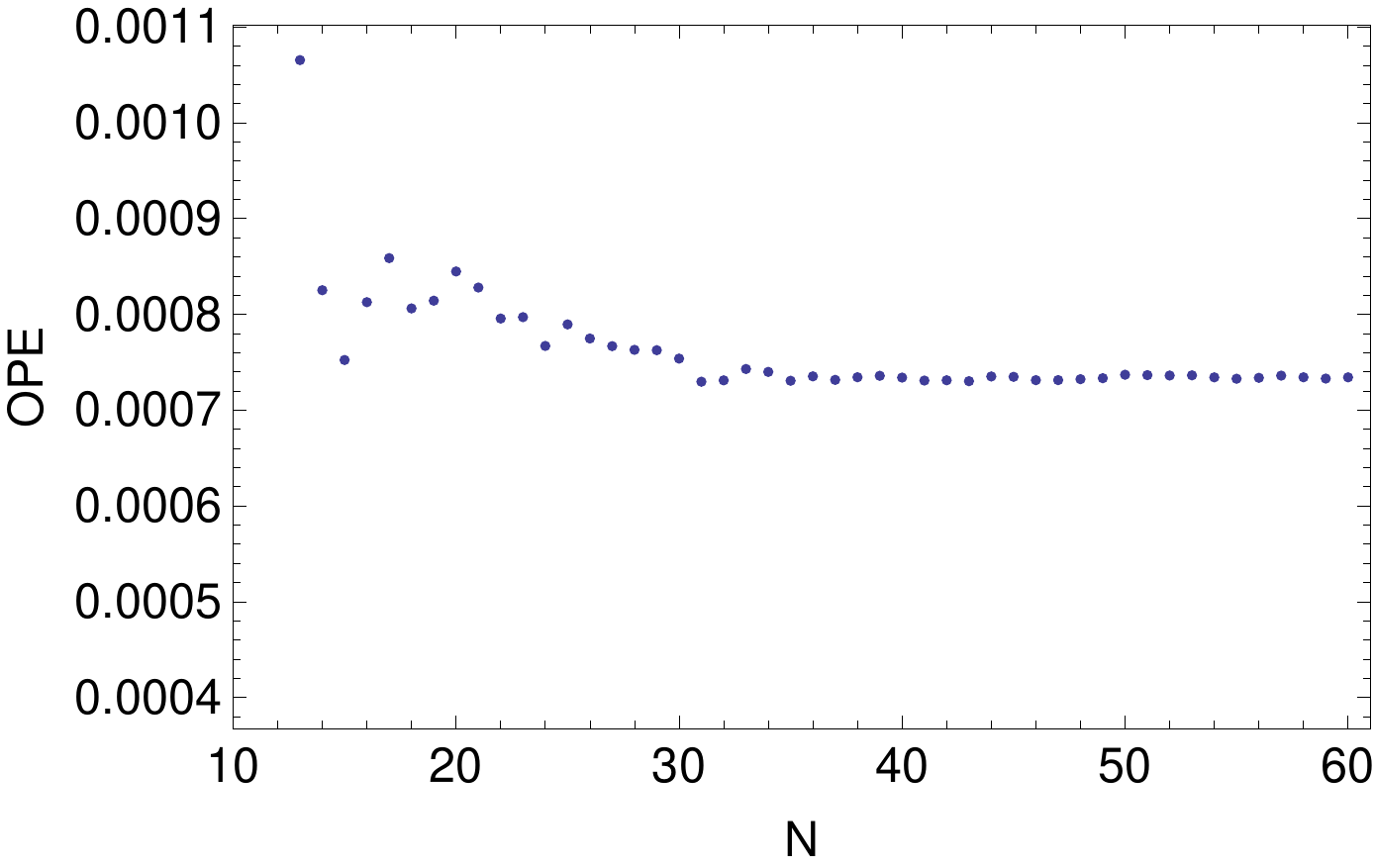}
\end{tabular}
\caption{Evolution of the OPE coefficients at the Ising point. We show only a subset of our results.}
\label{fig:OPEconv}
\end{figure}

\newpage 

\begin{table}[htbp]
  \centering
  \caption{Spectrum obtained with $N=60$. Below are operators with $L\leq 6$. The last column gives the OPE error estimate derived from examining OPE coefficient convergence. Recall that this estimate is derived by comparing  the variation of the OPE coefficient at the last jump, $\delta \lambda_{\mathcal O}\equiv |\lambda_{\mathcal O_{N=60}}-\lambda_{\mathcal O_{N=58}}|$, to the coefficient itself (see main text for more details). N/A indicates that the corresponding operator hasn't converged enough for an OPE error estimate to be available. }
  \vspace{0.5 cm}
    \begin{tabular}{|c|c|c|c|c|c|c|r|}
    \hline
    $L$     &  $\Delta_{\mbox{\tiny EFM}}$      & $\Delta$ &  Err$_{\Delta}$ (\%)       &  OPE$_{\mbox{\tiny EFM}}$     &   OPE    &   Err$_{\mbox{\tiny OPE}}$ (\%)    &  Err. Est. (\%) \bigstrut\\
    \hline
	\multirow{4}[8]{*}{0} &  1.000003 & 1     & 0.00025 & 0.4999997 & 0.5   & 6.98E-05 & \multicolumn{1}{c|}{1.1087E-05} \bigstrut\\
\cline{2-8}          & 4.0003 & 4     & 0.0076 & 1.56241E-02 & 0.015625 & 0.0059 & \multicolumn{1}{c|}{0.003} \bigstrut\\
\cline{2-8}          & 8.0817 & 8     & 1.0   & 2.17003E-04 & 0.00021973 & 1.2   & \multicolumn{1}{c|}{2.8} \bigstrut\\
\cline{2-8}          & 30.2000 & 29    & 4.1   & 2.46649E-07 & 0.0017688 & 100.0 & \multicolumn{1}{c|}{N/A} \bigstrut\\
    \hline
    \multirow{5}[10]{*}{2} & 2.0000 & 2     & 0     & 1.76777E-01 & 0.176777 & 0.0001 & \multicolumn{1}{c|}{0.00070} \bigstrut\\
\cline{2-8}          & 5.9979 & 6     & 0.035 & 2.61754E-03 & 0.00262039 & 0.1   & \multicolumn{1}{c|}{0.02} \bigstrut\\
\cline{2-8}          & 7.8600 & 6     & 31    & 8.66110E-05 & 0.00262039 & 96.7  & \multicolumn{1}{c|}{N/A} \bigstrut\\
\cline{2-8}          & 10.6200 & 11    & 3.5   & 4.11441E-05 & 9.6505E-06 & 326.3 & \multicolumn{1}{c|}{N/A} \bigstrut\\
\cline{2-8}          & 14.3267 & 14    & 2.3   & 8.60258E-07 & 1.9167E-06 & 55.1  & \multicolumn{1}{c|}{N/A} \bigstrut\\
    \hline
    \multirow{5}[10]{*}{4} & 4.0000 & 4     & 0     & 2.09627E-02 & 0.0209631 & 0.0021 & \multicolumn{1}{c|}{0.005} \bigstrut\\
\cline{2-8}          & 5.0003 & 5     & 0.0063 & 5.52411E-03 & 0.00552427 & 0.0030 & \multicolumn{1}{c|}{0.04} \bigstrut\\
\cline{2-8}          & 7.9920 & 8     & 0.1   & 4.63914E-04 & 0.00046138 & 0.5   & \multicolumn{1}{c|}{0.8} \bigstrut\\
\cline{2-8}          & 11.4067 & 12    & 4.9   & 1.26831E-05 & 1.0886E-05 & 16.5  & \multicolumn{1}{c|}{21.9} \bigstrut\\
\cline{2-8}          & 15.2600 & 16    & 4.6   & 2.07807E-06 & 4.0479E-07 & 413.4 & \multicolumn{1}{c|}{N/A} \bigstrut\\
    \hline
    \multirow{3}[6]{*}{6} & 6.0000 & 6     & 0     & 3.69140E-03 & 0.00369106 & 0.0092 & \multicolumn{1}{c|}{0.0006} \bigstrut\\
\cline{2-8}          & 6.9978 & 7     & 0.031 & 1.23528E-03 & 0.00123526 & 0.0013 & \multicolumn{1}{c|}{0.2} \bigstrut\\
\cline{2-8}          & 10.0009 & 10    & 0.0089 & 9.15865E-05 & 9.1798E-05 & 0.2   & \multicolumn{1}{c|}{2.3} \bigstrut\\
    \hline
    \end{tabular}%
  \label{tab:spec1}%
\end{table}%

\newpage

\begin{table}[htbp]
  \centering
  \caption{Spectrum obtained with $N=60$. Shown are operators with $L\geq 8$. The last column gives the OPE error estimate derived from examining OPE coefficient convergence. Recall that this estimate is derived by comparing  the variation of the OPE coefficient at the last jump, $\delta \lambda_{\mathcal O}\equiv |\lambda_{\mathcal O_{N=60}}-\lambda_{\mathcal O_{N=58}}|$, to the coefficient itself (see main text for more details).  N/A indicates that the corresponding operator hasn't converged enough for an OPE error estimate to be available. Highlighted rows correspond to operators selected by our criteria, namely OPE variation less than $15\%$ in the last two jumps.
  }
  \vspace{0.5 cm}
    \begin{tabular}{|c|c|c|c|c|c|c|r|}
    \hline
     $L$     &  $\Delta_{\mbox{\tiny EFM}}$      & $\Delta$ &  Err$_{\Delta}$ (\%)       &  OPE$_{\mbox{\tiny EFM}}$     &   OPE    &   Err$_{\mbox{\tiny OPE}}$ (\%)    &  Err. Est. (\%) \bigstrut\\
    \hline
    \multirow{4}[8]{*}{8} & 8.0000 & 8     & 0     & 7.34201E-04 & 7.34387E-04 & 2.53E-02 & \multicolumn{1}{c|}{0.2} \bigstrut\\
\cline{2-8}          & 9.0104 & 9     & 0.12  & 2.73456E-04 & 2.73566E-04 & 4.03E-02 & \multicolumn{1}{c|}{0.8} \bigstrut\\
\cline{2-8}          & 12.2231 & 12    & 1.9   & 1.79364E-05 & 1.93865E-05 & 7.5   & \multicolumn{1}{c|}{3.1} \bigstrut\\
\cline{2-8}          & 14.8267 & 16    & 7.3   & 1.65227E-06 & 5.02889E-07 & 228.6 & \multicolumn{1}{c|}{N/A} \bigstrut\\
    \hline
    \multirow{4}[8]{*}{10} & 10.0000 & 10    & 0.0   & 1.55108E-04 & 1.55092E-04 & 0.01  & \multicolumn{1}{c|}{1.8} \bigstrut\\
\cline{2-8}          & 10.9582 & 11    & 0.4   & 6.18464E-05 & 6.14425E-05 & 0.7   & \multicolumn{1}{c|}{2.0} \bigstrut\\
\cline{2-8}          & 13.0378 & 14    & 6.9   & 5.70613E-06 & 4.24073E-06 & 34.6  & \multicolumn{1}{c|}{40.1} \bigstrut\\
\cline{2-8}          & 15.6400 & 14    & 11.7  & 0     & 4.24073E-06 & 100.0 & \multicolumn{1}{c|}{N/A} \bigstrut\\
    \hline
    \multirow{3}[6]{*}{12} & 12.0000 & 12    & 0     & 3.40346E-05 & 3.39259E-05 & 0.3   & \multicolumn{1}{c|}{10.8} \bigstrut\\
\cline{2-8}          & 13.1573 & 13    & 1.2   & 1.33665E-05 & 1.39907E-05 & 4.5   & \multicolumn{1}{c|}{0.9} \bigstrut\\
\cline{2-8}          & 16.6882 & 16    & 4.3   & 6.92110E-07 & 9.49371E-07 & 27.1  & \multicolumn{1}{c|}{20.3} \bigstrut\\
    \hline
    \multirow{3}[6]{*}{14} & 14.0000 & 14    & 0.0   & 7.09873E-06 & 7.59497E-06 & 6.5   & \multicolumn{1}{c|}{32.3} \bigstrut\\
\cline{2-8}          & 14.5143 & 15    & 3.2   & 4.16397E-06 & 3.22191E-06 & 29.2  & \multicolumn{1}{c|}{5.2} \bigstrut\\
\cline{2-8}          & 17.1333 & 18    & 4.8   & 5.04814E-07 & 2.16042E-07 & 133.7 & \multicolumn{1}{c|}{69.9} \bigstrut\\
    \hline
    \multirow{2}[4]{*}{16} & 16.0069 & 16    & 0     & 1.93363E-06 & 1.72834E-06 & 11.9  & \multicolumn{1}{c|}{100} \bigstrut\\
\cline{2-8}          & 17.4200 & 17    & 2.5   & 6.73718E-07 & 7.48758E-07 & 10.0  & \multicolumn{1}{c|}{21.02} \bigstrut\\
    \hline
    18    & 18.0000 & 18    & 0     & 0     & 3.98122E-07 & 100.0 & \multicolumn{1}{c|}{N/A} \bigstrut\\
    \hline
    \multirow{2}[4]{*}{20} & 20.2069 & 20    & 1.0   & 1.84010E-07 & 9.25733E-08 & 98.8  & \multicolumn{1}{c|}{43.8} \bigstrut\\
\cline{2-8}          & 42.6333 & 29    & 47.0  & 0     & 8.15221E-04 & 100.0 & \multicolumn{1}{c|}{N/A} \bigstrut\\
    \hline
    \end{tabular}%
  \label{tab:spec2}%
\end{table}%

\newpage

\bibliography{Biblio}{}
\bibliographystyle{JHEP}

\end{document}